\documentclass[12pt,letterpaper]{article}
\usepackage[margin=1in]{geometry}
\usepackage[T1]{fontenc}
\usepackage{lmodern}
\usepackage{amsmath,amssymb,amsfonts,amsthm,mathtools}
\usepackage{graphicx,xcolor}
\usepackage{algorithm,algpseudocode}
\usepackage{booktabs,subcaption,float}
\usepackage{microtype}
\usepackage{natbib}
\usepackage{url}
\usepackage{setspace}
\usepackage{comment}
\usepackage[hyperfootnotes=false,hidelinks]{hyperref}
\usepackage{cleveref}
\bibpunct{(}{)}{;}{a}{,}{,}
\newtheorem{theorem}{Theorem}

\newtheorem{corollary}[theorem]{Corollary}
\theoremstyle{definition}
\newtheorem{definition}[theorem]{Definition}
\theoremstyle{remark}

\newenvironment{keywords}{\par\smallskip\noindent\textbf{Keywords: }}{\par\medskip}

\makeatletter
\@ifundefined{theHALG@line}
  {\newcommand{\theHALG@line}{\thealgorithm.\arabic{ALG@line}}}
  {\renewcommand{\theHALG@line}{\thealgorithm.\arabic{ALG@line}}}
\makeatother

\newcommand{\R}{\mathbb{R}}
\newcommand{\N}{\mathbb{N}}
\newcommand{\E}{\mathbb{E}}
\newcommand{\Pp}{\mathbb{P}}

\newcommand{\oas}{\operatorname{o}_{\rm a.s.}}

\newcommand{\ind}{\mathbf{1}}
\newcommand{\eps}{\varepsilon}

\newcommand{\argmax}{\operatorname*{arg\,max}}

\newcommand{\maybeincludegraphics}[2][]{%
  \IfFileExists{#2}{\includegraphics[#1]{#2}}{%
    \fbox{\begin{minipage}[c][0.20\textheight][c]{0.88\linewidth}\centering
    Figure file not included in this archive:\\ \texttt{\detokenize{#2}}
    \end{minipage}}}}

\begin{document}
\title{Asymptotic Anytime-Valid Quantile Inference under Local Differential Privacy}
\author{Leheng Cai$^1$$^3$\qquad Qirui Hu$^2$$^3$\thanks{Corresponding author.}\qquad Shuyuan Wu$^2$$^3$\\[0.6em]
\small $^1$Department of Statistics and Data Science, Tsinghua University\\
\small Beijing 100084, China\\[0.3em]
\small $^2$School of Statistics and Data Science,\\
\small Shanghai University of Finance and Economics, Shanghai 200433, China\\[0.4em]
\small $^3$All authors equally contributed to this article, and  are listed in the alphabetical order.\\
\small \texttt{cailh22@mails.tsinghua.edu.cn}, \texttt{huqirui@mail.shufe.edu.cn}\\
\small \texttt{wushuyuan@mail.sufe.edu.cn}}
\date{}
\maketitle

\begin{abstract}%
Sequential quantile inference is difficult under local differential privacy because every record is randomized before reaching the analyst and the limiting quantile variance depends on an unknown density. We develop an online procedure that combines randomized response with dynamically chained parallel stochastic gradient descent (P-SGD). The resulting Polyak--Ruppert estimator admits a strong Gaussian approximation. A cross-chain quadratic statistic, computed entirely from private iterates, consistently estimates the limiting variance without a separate online density estimator. These results yield asymptotic confidence sequences and, under polynomial chain growth, asymptotic time-uniform coverage. Arm-wise constructions support locally private quantile best-arm identification, time-uniform simple-regret bounds, and sequential A/B tests of quantile treatment effects. Simulations and salary-data analyses illustrate the finite-sample behavior and practical use of the proposed methods.
\end{abstract}

\begin{keywords}
anytime-valid inference, confidence sequences, local differential privacy, quantile estimation, stochastic approximation
\end{keywords}

\onehalfspacing
\section{Introduction}

Quantiles govern decisions in risk management, reliability, online experimentation, and resource allocation. Tail quantiles characterize rare delays and losses, while median and quantile effects describe distributional behavior that mean-based analyses can miss \citep{chen2008nonparametric,wang2012estimation,draghicescu2009quantile,chernozhukov2011inference,hu2022stochastic}. In streaming applications, these targets may be computed from salaries, educational records, device telemetry, or user interactions. Such records can remain sensitive even after conventional anonymization \citep{narayanan2006break,hua2016we,anand2019spearphone}. Local differential privacy (LDP) addresses this risk by requiring each participant to randomize a record before transmission, so the analyst never receives the raw value \citep{warner1965randomized,duchi2013local,erlingsson2014rappor}.

An anytime-valid LDP analysis of quantiles cannot be obtained by simply adding privacy noise to a classical estimator. User-side randomization contributes at the root-sample-size order and therefore changes the leading Gaussian variation \citep{cai2021cost}. At the same time, studentizing a quantile estimator requires the density at the target quantile. A non-private analyst can estimate this local feature from the empirical distribution, but query-dependent binary reports or stochastic gradients do not retain the raw observations needed to reconstruct it. A direct density estimate from separately privatized information would generally require an additional mechanism and privacy allocation. Consequently, a one-message-per-record estimator does not by itself provide the online scale estimate needed for inference.

Private quantile estimation has been studied under both central and local privacy. Central-model procedures include private sample quantiles, simultaneous quantiles, and algorithms for bounded domains \citep{dwork2009differential,smith2011privacy,gillenwater2021differentially,alabi2022bounded,ben2022archimedes}. Recent LDP work develops online self-normalized quantile inference, tuning-free distribution-function methods, and lightweight distributed protocols \citep{liu2023online,liu2024tuning,aamand2025lightweight}. These contributions provide point estimation or fixed-time inference, but they do not jointly provide online local privacy, time-uniform quantile inference, and a strongly consistent variance estimator formed solely from the stochastic-approximation iterates.

Time-uniform inference introduces a second  difficulty. A central limit theorem and a variance estimator that converges in probability can justify inference at a prespecified time, but they do not control repeated inspection or a data-dependent stopping time. Confidence sequences require coverage over the monitoring path \citep{robbins1970statistical,robbins1970boundary,howard2021time}; their asymptotic construction rests on an almost sure Gaussian coupling whose remainder is negligible at the time-uniform boundary scale, together with an almost surely consistent scale estimate \citep{waudby2024time}. For quantiles, the stochastic gradient is an indicator and changes discontinuously with the iterate. The smooth linearization and maximal arguments used for averaged stochastic approximation therefore do not directly deliver the required uniform coupling \citep{fang2018online,Lee2022,su2023higrad,li2022statistical,xie2024asymptotic}. For non-private quantiles, \citet{howard2022sequential} obtain nonasymptotic confidence sequences and sequential decision procedures, but the additional privatization and studentization problems remain under LDP.

Parallel stochastic-gradient recursions create variance replicates \citep{zhu2024high}, but consistency creates a third difficulty. With a fixed number of chains, cross-chain quadratic variation has no law-of-large-numbers stabilization. The chain count must diverge slowly enough that all but the newest chain have long trajectories, while the newest chain's contribution must be controlled separately. This balance makes initialization bias, nonlinear remainders, and stochastic-approximation error negligible in the aggregate. Approximation and variance bounds must therefore hold throughout catch-up, not only at balanced allocation times.
To address these issues, we propose a dynamically expanding system of parallel stochastic-gradient recursions.  Each record supplies one locally privatized binary gradient and is used only once. Pooling the chain-wise Polyak--Ruppert averages yields the quantile estimator, while a cross-chain quadratic statistic estimates its limiting variance entirely from the private iterates. A plateau-based allocation lets the number of chains increase while keeping all but the newest chain comparable in length; the newest chain is caught up within each plateau.
Under this construction, we establish a  strong Gaussian approximation, variance consistency, and the asymptotic time-uniform coverage needed for sequential decisions.

This article substantially extends the conference version \citep{cai2025timeuniform}. It replaces the conference version's chain-length-weighted variance statistic with the new estimator in \eqref{eq:variance-estimator} and proves an explicit convergence rate for the new statistic. It also establishes asymptotic time-uniform coverage, and develops locally private procedures and guarantees for quantile best-arm identification, time-uniform simple regret, and online quantile A/B testing, supported by new simulations and salary-data analyses.

The main contributions are as follows.
\begin{enumerate}
\item We construct a fully online LDP quantile estimator and variance estimator using \(O(\kappa_T)\) memory at time \(T\), where $\kappa_T$ is the number of chains. Both are post-processing of one private response per record, so variance estimation incurs no additional privacy loss.
\item We prove a strong Gaussian approximation with remainder \(o_{\rm a.s.}(T^{-1/2-\lambda})\) for an explicit range of \(\lambda>0\).
\item We establish   strong consistency, with an explicit rate, for the   variance estimator. This gives general asymptotic confidence sequences; a polynomial lower bound on chain growth further yields asymptotic time-uniform coverage.
\item We derive delayed-start guarantees for quantile best-arm identification and A/B testing and a time-uniform simple-regret bound.
\end{enumerate}

The rest of the paper is organized as follows. Section~\ref{sec:problem-method} introduces the setup and proposed method. Section~\ref{sec:theory} develops the Gaussian approximation, variance theory, and asymptotic confidence sequences. Section~\ref{sec:applications} gives the sequential decision procedures, followed by numerical experiments in Section~\ref{sec:experiments} and salary-data analyses in Section~\ref{sec:real-data}. Section~\ref{sec:discussion} concludes. Additional analyses are provided in Appendices~\ref{app:additional-numerics} and~\ref{app:gpa-real-data}.

\section{Methodology}\label{sec:problem-method}

\subsection{Privacy and Anytime-Valid Inference}

We begin with the privacy and sequential-inference notions used throughout; see \citet{dwork2006our,dwork2014algorithmic,duchi2013local} for broader treatments.

\begin{definition}[Differential Privacy]
A randomized algorithm \(\mathcal A\), taking a data set of individuals as input, is \((\eps,\delta)\)-differentially private if, for any neighboring data sets \(S\) and \(S'\) differing in one individual and any measurable event \(E\),
\[
\Pp\{\mathcal A(S)\in E\}\le e^\eps \Pp\{\mathcal A(S')\in E\}+\delta.
\]
When \(\delta=0\), the algorithm is \(\eps\)-differentially private.
\end{definition}

\begin{definition}[Local Differential Privacy]
A randomized map \(R:\mathcal X\to\mathcal Y\) is an
\((\eps,\delta)\)-local randomizer if it is an
\((\eps,\delta)\)-differentially private algorithm whose input is a single
record.
\end{definition}
 Compared with the
central model, LDP therefore imposes the stricter data-access and trust
requirement that the analyst never observes raw records.

For sequential inference, let \(\mathcal T\) be a totally ordered infinite time set. A sequence of intervals \((C_t)_{t\in\mathcal T}\) is a confidence sequence for a fixed parameter \(\theta\) if \(\Pp(\theta\in C_t\text{ for all }t\in\mathcal T)\ge 1-\alpha\). Throughout the paper, we use the notion of asymptotic confidence sequences in  \citet{waudby2024time}.

\begin{definition}[Asymptotic Confidence Sequence]
Let
\(C_t=[\widehat\theta_t-L_t,\widehat\theta_t+U_t]\),
where \(L_t,U_t>0\).  The intervals \((C_t)_{t\in\mathcal T}\)
form a \((1-\alpha)\)-asymptotic confidence sequence (AsympCS) for
\(\theta\) if there exists a (possibly unknown) non-asymptotic
\((1-\alpha)\)-confidence sequence
\[
C_t^\star=
[\widehat\theta_t-L_t^\star,\widehat\theta_t+U_t^\star],
\qquad t\in\mathcal T,
\]
centered at the same estimators, such that
\[
\Pp\!\left(\theta\in C_t^\star\text{ for every }t\in\mathcal T\right)
\ge1-\alpha,
\qquad
\frac{L_t^\star}{L_t}\longrightarrow1,
\quad
\frac{U_t^\star}{U_t}\longrightarrow1
\quad\text{almost surely}.
\]
\end{definition}
\begin{definition}[Asymptotic Time-Uniform Coverage]\label{def:time-uniform}
For each \(m\in\N\), let \((C_t(m))_{t\ge m}\) be a sequence of
random sets. This family has asymptotic time-uniform coverage at level
\(1-\alpha\) for \(\theta\) if
\[
\liminf_{m\to\infty}
\Pp\!\left(\theta\in C_t(m)\text{ for every }t\ge m\right)
\ge1-\alpha.
\]
\end{definition}

The two definitions describe different aspects of asymptotic validity. The AsympCS definition compares interval widths with those of a genuine confidence sequence, whereas Definition~\ref{def:time-uniform} concerns joint coverage after a growing monitoring start. For standard fixed Gaussian-mixture boundaries, sufficiently late monitoring can give coverage tending to one at different nominal levels. Recalibrating the boundary with the start time makes the delayed-start formulation sensitive to the nominal level; Section~\ref{sec:theory} establishes this sharper property for our construction. See \citet[Section~2.1]{waudby2024time} for the motivation behind the width-based definition.

\subsection{Locally Private Quantile Recursion}

Let \(\xi_1,\xi_2,\ldots\) be independent observations from a distribution \(F\) on \(\R\). Fix \(\tau\in(0,1)\) and write
\[
q_\tau=\inf\{x:F(x)\ge \tau\}
\]
for the target quantile. Throughout the main theory, we assume that \(F(q_\tau)=\tau\) and that the density is positive at \(q_\tau\).

We use the binary-response mechanism of \citet{liu2023online}. Independently of
\(\xi\), draw \(W\sim\operatorname{Bernoulli}(r)\) and
\(V\sim\operatorname{Bernoulli}(1/2)\). The parameter \(r\in(0,1]\) is the
truthful-response rate, with \(r=1\) giving the non-private benchmark. For
\(\zeta=(\xi,W,V)\) and a query point \(x\), define
\begin{align}\label{eq:private-gradient}
G(x,\zeta)
&=\frac{1+r-2r\tau}{2}\{\ind(\xi\le x)W+(1-W)(1-V)\}  \\
&\quad -\frac{1-r+2r\tau}{2}\{\ind(\xi>x)W+(1-W)V\}. \notag
\end{align}
The mechanism is \(\eps\)-LDP with
$
\eps=\log(1+r)-\log(1-r)
$
for \(r<1\). This guarantee holds conditionally on the previous public transcript when the query point is chosen from that transcript. Each arriving record is queried only once. Direct calculation gives
\begin{equation}\label{eq:g-def}
g(x):=\E\{G(x,\zeta)\}=r\{F(x)-F(q_\tau)\}=r\{F(x)-\tau\}.
\end{equation}
Hence \(G(x,\zeta)\) is a locally private stochastic gradient for the quantile estimating equation \(F(x)-\tau=0\). A single-chain recursion is
\begin{equation}\label{eq:single-sgd}
x_{t+1}=x_t-\eta_{t+1}G(x_t,\zeta_{t+1}),\qquad t\ge0.
\end{equation}
Polyak--Ruppert averaging of \eqref{eq:single-sgd} gives a consistent quantile estimator. Inference also requires its limiting variance, which involves the unknown density \(f(q_\tau)\). The parallel construction below estimates this scale directly from the private trajectories.

\subsection{Dynamically Chained Parallel SGD}

To estimate the asymptotic variance from the iterates themselves, we run many independent SGD chains with identical initial values. At time \(T\), let \(\kappa_T\) be the number of active chains and let \(T_k\) be the number of observations assigned to chain \(k\). Chain \(k\) evolves as
\[
x_{k,t+1}=x_{k,t}-\eta_{t+1}G(x_{k,t},\zeta_{k,t+1}),
\qquad t=0,\ldots,T_k-1,
\]
with common fixed initialization \(x_{k,0}=x_0\). The deterministic allocation preserves independence across chains.

The number of chains must diverge for the quadratic variance estimator to be consistent, but it must grow slowly enough that all but the newest chain have long trajectories. Let \(h:\N\to\N\) be nondecreasing and piecewise constant, set \(\kappa_T=h(T)\), and require unit jumps, \(h(T+1)-h(T)\in\{0,1\}\). Write \(K_0=h(1)\ge2\) and
\[
m_j=\bigl|\{T:h(T)=K_0+j\}\bigr|,\qquad j\ge0,
\]
for the plateau lengths. We impose
\begin{equation}\label{eq:plateau-schedule}
m_0\ge K_0,
\qquad
m_j\ge \frac{1}{K_0+j-1}\sum_{i=0}^{j-1}m_i,
\qquad j\ge1.
\end{equation}
We use shortest-chain allocation, with ties broken by the smallest index. Condition~\eqref{eq:plateau-schedule} gives a newly introduced chain enough time to catch up before another chain is introduced. Thus, at every time, either all active chains differ in length by at most one, or only the newest chain is still catching up while the other chains differ in length by at most one. For example, for constants \(b>1\) and \(c>0\),
\[
h(T)=\lfloor c\log_b T\rfloor+K_0,
\qquad
b^{1/c}>\max\{K_0^{-1}+2,K_0\},
\]
satisfies \eqref{eq:plateau-schedule}. Such logarithmic schedules meet the upper-growth conditions used for the Gaussian approximation and variance consistency. The sharp delayed-start result below additionally requires polynomial lower growth, for which a plateau-adjusted polynomial schedule may be used.

We define the plateau-adjusted version of a target unit-jump
schedule recursively. Starting from \(h(1)=K_0\), a proposed increase from
\(K_0+j\) to \(K_0+j+1\) is accepted only after the current plateau has met
the corresponding lower bound in \eqref{eq:plateau-schedule}; otherwise the
increase is postponed and \(h\) is held constant. For the logarithmic and
polynomial target schedules used below, this adjustment preserves their
stated growth orders.

Algorithm~\ref{alg:allocation} gives a one-step state update that is invoked
when each observation arrives.

\begin{algorithm}[htbp]
\caption{Online Balanced Dynamic Allocation for Parallel Runs}\label{alg:allocation}
\begin{algorithmic}[1]
\State \textbf{Input at arrival \(t\):} nondecreasing chain-count function
\(h(\cdot)\) and the persistent count vector \({\tt nums}\) after arrivals
\(1,\ldots,t-1\), with \({\tt nums}=()\) when \(t=1\)
\While{\(|{\tt nums}|<h(t)\)}
        \State Append \(0\) to \({\tt nums}\)
\EndWhile
\State \(k_t\leftarrow\) first index attaining \(\min_j {\tt nums}[j]\)
\State \({\tt nums}[k_t]\leftarrow {\tt nums}[k_t]+1\)
\State \textbf{Output:} \(k_t\) and the updated persistent vector \({\tt nums}\)
\State \textbf{Memory:} \(|{\tt nums}|=h(t)=\kappa_t\)
\end{algorithmic}
\end{algorithm}

The online quantile estimator is the weighted average of all chain-wise Polyak--Ruppert averages,
\[
\widehat q_T=\frac{1}{T}\sum_{k=1}^{\kappa_T}\sum_{t=1}^{T_k}x_{k,t}.
\]

For \(T\ge K_0\), define
\begin{equation}\label{eq:variance-estimator}
\widehat\sigma_T^2=\frac{1}{\kappa_T}
\sum_{k=1}^{\kappa_T}
\left\{\frac{1}{\sqrt{T_k }}
\sum_{t=1}^{T_k }\bigl(x_{k,t}-\widehat q_T\bigr)\right\}^2.
\end{equation}

Write \(\widehat\sigma_T=(\widehat\sigma_T^2)^{1/2}\). For interval construction, we use the positive scale \(\widetilde\sigma_T=\max\{\widehat\sigma_T,T^{-1}\}\). The vanishing floor gives positive widths even when the early chain averages coincide; variance consistency makes it inactive eventually almost surely.

Writing \(\bar x_{k,T}=T_k^{-1}\sum_{t=1}^{T_k}x_{k,t}\), the statistic in \eqref{eq:variance-estimator} averages \(T_k(\bar x_{k,T}-\widehat q_T)^2\) equally across chains. These trajectory-level replicates stabilize the variance estimate as the number of chains grows, whereas the quantile estimator weights chains by their sample sizes. Whenever \(\widehat\sigma_T>0\), the corresponding plug-in density estimator is
\[
\widehat f_T(q_\tau)=
\frac{\{1-r^2(2\tau-1)^2\}^{1/2}}{2r\widehat\sigma_T}.
\]
The estimator \(\widehat q_T\), the variance estimator \(\widehat\sigma_T^2\), this density estimator, and all downstream intervals are post-processing of locally privatized responses. They therefore inherit the LDP guarantee of \eqref{eq:private-gradient}.

\section{Time-Uniform Approximation and Inference}\label{sec:theory}

We state the assumptions used in the asymptotic analysis.
\begin{description}
\item[(A1)] The distribution \(F\) is absolutely continuous on \(\R\) with density \(f\), and \(f(q_\tau)>0\).
\item[(A2)] The density \(f\) is bounded and globally Lipschitz.
\item[(A3)] The step size satisfies \(\eta_s=\eta_0s^{-a}\), \(s\ge1\), for some constants \(\eta_0>0\) and \(a\in(1/2,1)\).
\item[(A4)] The function \(h\) has unit jumps and satisfies \eqref{eq:plateau-schedule}. For some \(q>0\) and \(\nu\in(0,1-1/(2a))\),
\[
\kappa_T\to\infty,
\qquad
\log T=O(\kappa_T^q),
\qquad\text{and}\qquad
\kappa_T=O\bigl(T^{1-1/(2a)-\nu}\bigr).
\]

\item[(A5)] For some constants
\(c_\kappa>0\) and
\(\beta\in\bigl(0,1-1/(2a)-\nu\bigr]\),
\[
\kappa_T\ge c_\kappa T^\beta
\qquad\text{for all sufficiently large }T.
\]
\end{description}
Assumptions (A1)--(A2) impose positivity at the target quantile and global Lipschitz regularity of the density. Assumption (A3) is the standard decreasing-step condition for Polyak--Ruppert averaging. Assumption (A4) balances two requirements: the growing collection of independent chains supports cross-chain variance estimation, while the upper growth bound leaves all but the newest chain sufficiently long.   Assumption (A5), imposed only when stated, strengthens variance convergence enough to obtain asymptotic time-uniform coverage and the sequential decision guarantees built from it.

\begin{theorem}[Strong Gaussian Approximation]\label{thm:gaussian-approx}
Under Assumptions (A1)--(A4), put
\[
b=\frac{1}{2a}+\nu,
\qquad
\lambda_q=\min\left\{b-\frac12,\ a\nu,\
\frac{b(1-a)}2,\ \frac{ba}{4}\right\}>0.
\]
There exist i.i.d. Gaussian random variables \(Z_1,Z_2,\ldots\) with mean zero and variance
\begin{equation}\label{eq:sigma2}
\sigma^2=\frac{1-r^2(2\tau-1)^2}{4r^2f^2(q_\tau)}
\end{equation}
such that, for every \(0<\lambda<\lambda_q\),
\[
\left|\widehat q_T-q_\tau-\frac1T\sum_{i=1}^T Z_i\right|
=\oas\bigl(T^{-1/2-\lambda}\bigr).
\]

\end{theorem}
Theorem~\ref{thm:gaussian-approx} couples the entire estimation path with one sequence of independent Gaussian increments. Although the iterates within a chain are dependent, their pooled average has an almost-sure approximation error that is polynomially smaller than \(T^{-1/2}\), and hence negligible at the law-of-the-iterated-logarithm scale used for time-uniform inference. The result applies throughout chain catch-up as well as at balanced allocation times.

The mean-square Gaussian approximation of \citet{zhu2024high} supports fixed-time inference from parallel runs; the present argument additionally controls approximation error along the monitoring path. Almost-sure couplings for averaged stochastic approximation are developed by \citet{xie2024asymptotic} under smoothness conditions on the stochastic gradient. For the quantile score, boundedness of the density instead gives \(\|\ind(\xi\le x)-\ind(\xi\le y)\|_2\lesssim |x-y|^{1/2}\). This weaker continuity requires a separate martingale bound for indicator changes, which accounts for the \(ba/4\) term in \(\lambda_q\).

The privacy budget changes the first-order scale, not merely a higher-order
remainder. Since
\(r=(e^\eps-1)/(e^\eps+1)=\tanh(\eps/2)\), the variance inflation relative to
the \(r=1\) recursion is
\[
\frac{\sigma^2(r)}{\sigma^2(1)}
=\frac{r^{-2}-(2\tau-1)^2}{1-(2\tau-1)^2}.
\]
It increases as privacy becomes stronger, i.e., as \(r\) (equivalently \(\epsilon\)) decreases. It also equals \(r^{-2}\) at the median,
and gives \(\sigma(r)\sim\{\eps f(q_\tau)\}^{-1}\) as \(\eps\downarrow0\).
Thus stronger privacy widens both fixed-time intervals and confidence
sequences at first order; increasing the number of chains stabilizes
studentization but cannot remove this information loss.

At \(r=1\), \eqref{eq:private-gradient} reduces to the ordinary quantile
score and \eqref{eq:sigma2} becomes the classical variance
\(\tau(1-\tau)/f^2(q_\tau)\).
At $r=1$, however, the present procedure does not reduce to the empirical-quantile construction of \citet{howard2022sequential}.
Their non-private intervals are non-asymptotic and can be
uniform over both time and quantile level; ours remain fixed-quantile
stochastic-approximation intervals with asymptotic validity.

The next theorem establishes strong consistency of \(\widehat\sigma_T^2\) and separates the errors due to cross-chain replication and finite chain lengths.
\begin{theorem}[Variance Consistency]\label{thm:variance-consistency}
Under Assumptions (A1)--(A4), let
\[
c_a=\min\left\{a-\frac12,\frac{1-a}{2},\frac a4,\frac12\right\}.
\]
For every \(0<c<c_a\) and \(0<\xi<c\), the
estimator in \eqref{eq:variance-estimator} satisfies, as \(T\to\infty\),
\begin{equation}\label{eq:variance-general-rate}
\left|\widehat\sigma_T^2-\sigma^2\right|
=O_{\rm a.s.}\left\{
\sqrt{\frac{\log\kappa_T}{\kappa_T}}
+\left(\frac{T}{\kappa_T}\right)^{-c+\xi}
\right\}.
\end{equation}
If (A5) also holds, then, for every \(
0<\zeta<\zeta_\sigma
:=\min\left\{\beta/2,\left(1/(2a)+\nu\right)c_a\right\}\),
\begin{equation}\label{eq:variance-polynomial-rate}
\left|\widehat\sigma_T^2-\sigma^2\right|
=\oas(T^{-\zeta}).
\end{equation}
\end{theorem}

The first term in \eqref{eq:variance-general-rate} is the cross-chain averaging error; the second is the within-chain stochastic-approximation remainder. Thus consistency requires both \(\kappa_T\to\infty\) and \(T/\kappa_T\to\infty\).
A further consequence is strong consistency of the plug-in density estimator \(\widehat f_T(q_\tau)\) defined in Section~\ref{sec:problem-method}. Since \(\sigma^2>0\), its denominator is positive eventually almost surely, and continuous mapping gives \(\widehat f_T(q_\tau)\to f(q_\tau)\).

Let \(\gamma_{T,m}\) be any deterministic symmetric two-sided confidence-sequence half-width for the mean of i.i.d. \(N(0,1)\) variables, started at time \(m\). That is,
\[
\Pp\left(\forall T\ge m:\left|T^{-1}\sum_{i=1}^T \widetilde Z_i\right|\le \gamma_{T,m}\right)\ge 1-\alpha
\]
for i.i.d.\ standard normal \(\widetilde Z_i\).

\begin{theorem}[Asymptotic Confidence Sequence]\label{thm:asympcs}
Under Assumptions (A1)--(A4), the intervals
\[
C_T=\left[\widehat q_T-\widetilde\sigma_T\gamma_{T,m},\,
          \widehat q_T+\widetilde\sigma_T\gamma_{T,m}\right],\qquad T\ge m\ge K_0,
\]
form a \((1-\alpha)\)-asymptotic confidence sequence for \(q_\tau\).
\end{theorem}

Theorem \ref{thm:asympcs} provides a general framework for constructing AsympCSs for quantiles under the LDP setting, requiring only a confidence sequence for Gaussian random variables with unit variance.
Existing confidence sequences for Gaussian variables in the literature include two useful choices of \(\gamma_{T,m}\): the stitched Gaussian boundary of \citet{howard2021time},
\begin{equation}\label{eq:stitched-boundary}
\gamma_{T,m}=1.7\sqrt{\frac{\log\log\{\max(2T/m,e)\}+0.72\log(10.4/\alpha)}{T}},
\end{equation}
and the normal-mixture boundary of \citet{robbins1970statistical,robbins1970boundary},
\begin{equation}\label{eq:mixture-boundary}
\gamma_{T,1}=\sqrt{\frac{2(T\rho^2+1)}{T^2\rho^2}\log\left(\frac{\sqrt{T\rho^2+1}}{\alpha}\right)},\qquad \rho>0.
\end{equation}
The stitched boundary has the LIL order \(\sqrt{\log\log T/T}\). The mixture boundary has order \(\sqrt{\log T/T}\) but is often narrower for the sample sizes used in practice after \(\rho\) is tuned to the relevant horizon.

For the asymptotic time-uniform coverage result, take \(0<\alpha\le1/2\); the mixture
parameter must depend on the monitoring start. Let
\[
c_\alpha=-2\log\alpha+\log(-2\log\alpha)+1,
\qquad
\widetilde\rho_{m,\alpha}^2=\frac{c_\alpha}{m\log(m\vee e)},
\]
where \(c_\alpha>0\), and define
\begin{equation}\label{eq:delayed-mixture-boundary}
\gamma^{\rm GM}_{T,m}(\alpha)
=\sqrt{\frac{2(T\widetilde\rho_{m,\alpha}^2+1)}
{T^2\widetilde\rho_{m,\alpha}^2}
\log\left(\frac{\sqrt{T\widetilde\rho_{m,\alpha}^2+1}}{\alpha}\right)},
\qquad T\ge m.
\end{equation}

\begin{corollary}[Sharp Coverage of the Gaussian-Mixture AsympCS]
\label{cor:gm-sharp-coverage}
Assume (A1)--(A5). The delayed-start intervals
\begin{equation}\label{eq:gm-sharp-asympcs}
C_T^{\rm GM}(m)=
\left[\widehat q_T-\widetilde\sigma_T\gamma^{\rm GM}_{T,m}(\alpha),\,
\widehat q_T+\widetilde\sigma_T\gamma^{\rm GM}_{T,m}(\alpha)\right],
\qquad T\ge m,
\end{equation}
have sharp asymptotic time-uniform \((1-\alpha)\)-coverage:
\begin{equation}\label{eq:gm-sharp-coverage}
\lim_{m\to\infty}
\Pp\left(\forall T\ge m:\ q_\tau\in C_T^{\rm GM}(m)\right)
=1-\alpha.
\end{equation}
\end{corollary}

A pointwise confidence interval follows immediately.
\begin{corollary}[Fixed-Time Confidence Interval]\label{cor:pointwise-ci}
Under (A1)--(A4), the confidence interval
\[
\left[\widehat q_T-\widetilde\sigma_T z_{1-\alpha/2}/\sqrt{T},\,
      \widehat q_T+\widetilde\sigma_T z_{1-\alpha/2}/\sqrt{T}\right]
\]
  has coverage converging to \(1-\alpha\) for \(q_\tau\) as
\(T\to\infty\), where \(z_{1-\alpha/2}\) is the
\((1-\alpha/2)\)-standard normal quantile.
\end{corollary}

Algorithm~\ref{alg:ldp-asympcs} summarizes the online implementation. All updates count toward \(T\) and the local step-size indices; the monitoring start \(m\) delays reporting without resetting either clock.

\begin{algorithm}[htbp]
\caption{Online LDP Asymptotic Confidence Sequence for a Quantile}
\label{alg:ldp-asympcs}
\begin{algorithmic}[1]
\State \textbf{Input:} open-ended stream \(\xi_1,\xi_2,\ldots\), level
\(\tau\), truthful-response rate \(r\), confidence level \(1-\alpha\),
chain function \(h\) satisfying \eqref{eq:plateau-schedule}, step sizes
\(\eta_s\), initial value \(x_0\), start time \(m\ge K_0\)
\State Initialize \(t\leftarrow0\), \(\kappa\leftarrow0\), and the persistent
count vector \((n_1,\ldots,n_\kappa)\leftarrow()\)
\For{each arriving observation \(\xi\)}
    \State \(t\leftarrow t+1\); \quad \(\xi_t\leftarrow\xi\); \quad
    \(\kappa_{\rm old}\leftarrow\kappa\)
    \State Invoke Algorithm~\ref{alg:allocation} with
    \((t,h,(n_1,\ldots,n_\kappa))\), obtaining selected chain \(k\) and
    updated counts \((n_1,\ldots,n_{h(t)})\)
    \State \(\kappa\leftarrow h(t)\)
    \If{\(\kappa>\kappa_{\rm old}\)}
        \For{\(j=\kappa_{\rm old}+1,\ldots,\kappa\)}
            \State Initialize \(x_j\leftarrow x_0\) and \(\bar x_j\leftarrow0\)
        \EndFor
    \EndIf
    \State \(s\leftarrow n_k\) \Comment{local sample size after allocation}
    \State Obtain the private response \(G(x_k,\zeta_t)\) from
    \eqref{eq:private-gradient}
    \State \(x_k\leftarrow x_k-\eta_sG(x_k,\zeta_t)\)
    \State \(\bar x_k\leftarrow\{(s-1)\bar x_k+x_k\}/s\)
    \State \(\widehat q_t\leftarrow\sum_{j=1}^{\kappa}(n_j/t)\bar x_j\)
    \If{\(\min_{j\le\kappa}n_j\ge1\)}
        \State \(\widehat\sigma_t^2\leftarrow
        \kappa^{-1}\sum_{j=1}^{\kappa}n_j
        (\bar x_j-\widehat q_t)^2\)
        \If{\(t\ge m\)}
            \State \(\widetilde\sigma_t\leftarrow\max\{\widehat\sigma_t,t^{-1}\}\), \quad \(s_t\leftarrow\widetilde\sigma_t\gamma_{t,m}\); report \(\widehat q_t\) and
            \([\widehat q_t-s_t,\widehat q_t+s_t]\)
        \EndIf
    \EndIf
\EndFor
\end{algorithmic}
\end{algorithm}
\section{Sequential Applications}\label{sec:applications}

 This section shows how the proposed asymptotic confidence sequences support  two standard sequential tasks. The first is quantile best-arm identification, where the analyst adaptively samples arms and seeks to identify an optimal arm. The second is A/B testing, where the analyst monitors a quantile treatment effect without inflating the false-positive probability.
 In both cases, privacy is inherited from the arm-wise LDP randomizers, and validity follows from the  asymptotic time-uniform coverage.

\subsection{Locally Private Quantile Best-Arm Identification}\label{subsec:qlucb}

Consider a $K$-armed bandit problem with arms
$\{1,\dots,K\}$, where each arm $k$ is associated with an unknown reward distribution $F_k$.
Each pull of arm $k$ supplies a fresh independent sample from $F_k$; one LUCB
round below contains two such pulls.

Let $\tau\in(0,1)$ be a fixed quantile level, and denote by $Q_k$ the quantile function of arm $k$.
We assume that each $F_k$ is continuous at $Q_k(\tau)$ and admits a positive density at that point, i.e.,
$f_k\!\big(Q_k(\tau)\big)>0$.
Under this condition, the $\tau$-quantile of each arm is well defined and unique.
For the guarantees below, we further assume that every arm
satisfies the arm-wise versions of Assumptions (A1)--(A4).  The asymptotic time-uniform guarantees additionally require the lower chain-growth condition
(A5).
The objective is to stop as early as possible and return an $\epsilon$-optimal arm (defined below). 

We say that arm \(k\) is
\(\epsilon\)-optimal if its target quantile is within \(\epsilon\) of the
largest arm quantile. This differs from the quantile-level perturbation used by
\citet{howard2022sequential} and is natural here because every arm-wise
estimator targets the fixed quantile \(Q_k(\tau)\). Writing
\(q_\star=\max_jQ_j(\tau)\), define
\[
\mathcal A_\epsilon
=\{k\in[K]:q_\star-Q_k(\tau)\le\epsilon\}.
\]
The Quantile-LUCB algorithm proceeds in rounds, maintaining lower and upper confidence bounds for each arm and sampling the current leader and its strongest challenger.
At the beginning of round \(t\), let \(N_{k,t}\) denote the number of observations received from arm \(k\).

For a monitoring start \(m\), let
\(C_{k,n}^{\tau,m}=[L_{k,n}^{\tau},U_{k,n}^{\tau}]\), \(n\ge m\), be the arm-\(k\) delayed-start Gaussian-mixture AsympCS in
\eqref{eq:gm-sharp-asympcs}, constructed at error level \(\delta/K\). Thus, with \(\widehat Q_{k,n}(\tau)\) denoting the estimate based on \(n\) observations from arm \(k\),
\[
L_{k,n}^{\tau}=\widehat Q_{k,n}(\tau)-l_{k,n},
\qquad
U_{k,n}^{\tau}=\widehat Q_{k,n}(\tau)+u_{k,n}.
\]
At global round \(t\), the algorithm uses these endpoints at local time \(N_{k,t}\). Define the joint monitoring event
\[
\mathcal E_m
=
\bigcap_{k=1}^K\bigcap_{n\ge m}
\left\{Q_k(\tau)\in[L_{k,n}^{\tau},U_{k,n}^{\tau}]\right\}.
\]
Corollary~\ref{cor:gm-sharp-coverage} applied arm by arm and a
union bound give
\[
\liminf_{m\to\infty}\Pp(\mathcal E_m)\ge1-\delta.
\]
The delayed Gaussian-mixture boundary tends to zero at each fixed
\(m\).  Variance consistency then gives the almost-sure shrinking-width event
\[
\mathcal W
=\bigcap_{k=1}^K
\left\{U_{k,n}^{\tau}-L_{k,n}^{\tau}\to 0\right\},
\quad \Pp(\mathcal W)=1.
\]

\begin{algorithm}[htbp]
\caption{Quantile-LUCB}\label{alg:LUCB}
\begin{algorithmic}[1]
\State \textbf{Input:} target quantile \(\tau\in(0,1)\), approximation parameter
\(\epsilon\ge0\), confidence level \(\delta\in(0,1)\), monitoring start \(m\ge K_0\)
\State Sample each arm \(m\) times; set \(N_{k,0}\leftarrow m\) for all \(k\in[K]\) and \(t\leftarrow0\)
\While{\(L_{k,N_{k,t}}^{\tau}+\epsilon/2<\max_{j\ne k}U_{j,N_{j,t}}^{\tau}-\epsilon/2\) for every \(k\in[K]\)}
    \State Choose \(h_t\in\argmax_{k\in[K]}L_{k,N_{k,t}}^{\tau}\)
    \State Choose \(\ell_t\in\argmax_{j\ne h_t}U_{j,N_{j,t}}^{\tau}\)
    \State Sample arms \(h_t\) and \(\ell_t\)
    \State Set \(N_{k,t+1}\leftarrow N_{k,t}+\ind\{k\in\{h_t,\ell_t\}\}\) for every \(k\in[K]\)
    \State Update the local estimators and endpoints at sample sizes \(N_{k,t+1}\)
    \State \(t\leftarrow t+1\)
\EndWhile
\State \textbf{Output:} any arm \(k\) such that
\(L_{k,N_{k,t}}^{\tau}+\epsilon/2\ge\max_{j\ne k}U_{j,N_{j,t}}^{\tau}-\epsilon/2\)
\end{algorithmic}
\end{algorithm}

The strict population margin associated with this stopping rule is
\[
\Gamma_\epsilon
=
\max_{k\in[K]}
\left\{Q_k(\tau)+\epsilon/2-
\max_{j\ne k}\bigl[Q_j(\tau)-\epsilon/2\bigr]\right\}.
\]
The margin is positive automatically when \(\epsilon>0\).
When \(\epsilon=0\), it is positive exactly when the best arm is unique.

\begin{theorem}[{{Quantile-LUCB Validity}}]
\label{thm:QLUCB}
Suppose every arm satisfies the arm-wise versions of Assumptions
(A1)--(A5) and \(K\ge2\) is fixed.  For each integer \(m\ge K_0\), let
\(\pi_m\in\N\cup\{\infty\}\) be the number of post-initialization LUCB
rounds, and let \(\widehat k_m\) denote the output on
\(\{\pi_m<\infty\}\).  If \(\Gamma_\epsilon>0\), then
\begin{equation}\label{eq:qlucb-validity}
\liminf_{m\to\infty}
\Pp\!\left(
\pi_m<\infty,\ \widehat k_m\in\mathcal A_\epsilon
\right)
\ge1-\delta.
\end{equation}
\end{theorem}

Theorem~\ref{thm:QLUCB} guarantees eventual stopping and asymptotically  \(\epsilon\)-optimal selection  as $m\to\infty$. The positive margin covers both approximate selection with tied best arms and exact selection with a unique best arm. The non-private analysis of \citet{howard2022sequential} additionally gives finite-sample pull bounds from nonasymptotic quantile confidence sequences, using its quantile-level approximation criterion.

\subsection{Time-Uniform Simple-Regret Bound}\label{subsec:time-uniform-regret}

Beyond the stopping guarantee in Theorem \ref{thm:QLUCB}, the arm-wise AsympCSs also
provide a time-uniform guarantee for the simple regret of the current
recommendation. For this purpose, we let the sampling and recommendation
process continue after a stopping criterion could first be met, so that it is defined at every monitoring time. Arm choices are predictable
from the public transcript, and each selected arm supplies a fresh independent
reward before local privatization.

We initialize every arm with $m$ observations and let $t \ge 0$ denote the
number of subsequent arm pulls. Write $q_k = Q_k(\tau)$,
$q^\ast = \max_{k \in [K]} q_k$,
and
\[
\sigma_k^2
=
\frac{1-r^2(2\tau-1)^2}
{4r^2 f_k^2\left\{Q_k(\tau)\right\}}.
\]
At global pull count $t$, let $\widehat{k}_t
\in
\arg\max_{k \in [K]} L_{k,N_{k,t}}$, $j_t
\in
\arg\max_{j \in [K]} U_{j,N_{j,t}}$,
and define the simple regret of the current recommendation and its observable
upper bound by
\[
S_t
=
q^\ast-q_{\widehat{k}_t},
\qquad
B_t
=
U_{j_t,N_{j_t,t}}
-
L_{\widehat{k}_t,N_{\widehat{k}_t,t}}.
\]

\begin{theorem}[Time-Uniform Simple-Regret Bound]\label{thm:time-uniform-regret}
Suppose every arm satisfies the arm-wise versions of Assumptions
(A1)--(A5) and $K \ge 2$ is fixed. For each integer $m \ge K_0$, let
$[L_{k,n},U_{k,n}]$, $n \ge m$, be the arm-$k$
Gaussian-mixture AsympCS in \eqref{eq:gm-sharp-asympcs}, constructed at error level
$\delta/K$. Then
\begin{equation}\label{eq:local-count-regret-bound}
\liminf_{m\to\infty}
\mathbb{P}
\left(
\forall t \ge 0:
S_t
\le
B_t
\le
U_{j_t,N_{j_t,t}}
-
L_{j_t,N_{j_t,t}}
\right)
\ge
1-\delta .
\end{equation}
\end{theorem}

On the joint coverage event, $q^\ast
\le
U_{j_t,N_{j_t,t}}$ and $q_{\widehat{k}_t}
\ge
L_{\widehat{k}_t,N_{\widehat{k}_t,t}}$,
which yields $S_t \le B_t$. Moreover, since $\widehat{k}_t$ maximizes the
lower confidence bound,
\(
L_{\widehat{k}_t,N_{\widehat{k}_t,t}}
\ge
L_{j_t,N_{j_t,t}},
\)
and hence
\[
B_t
\le
U_{j_t,N_{j_t,t}}
-
L_{j_t,N_{j_t,t}}.
\]
Thus \(B_t\) is at most the width of the confidence sequence for
the arm whose upper confidence bound currently challenges the recommendation.

More precisely, for fixed $m$, $K$, and $\delta$, whenever
$N_{j_t,t}\to\infty$,
\[
B_t
\le
2\widetilde{\sigma}_{j_t,N_{j_t,t}}
\gamma^{\mathrm{GM}}_{N_{j_t,t},m}(\delta/K)
=
\mathcal{O}_{\mathrm{a.s.}}
\left\{
\sigma_{j_t}
\sqrt{
\frac{\log N_{j_t,t}}
{N_{j_t,t}}
}
\right\}.
\]
Therefore, under a general predictable allocation, the relevant effective
sample size is the local sample size of the arm that continues to challenge
the current recommendation.

A global-time rate follows under forced exploration. If, for some constant
$c_0>0$,
\(
N_{k,t}
\ge
m+ {c_0t}/{K}
\)
eventually for every $k\in[K]$, then, for fixed $m$, $K$, and $\delta$,
\[
B_t
=
\mathcal{O}_{\mathrm{a.s.}}
\left\{
\overline{\sigma}
\sqrt{
\frac{K\log(t\vee e)}
{c_0t}
}
\right\},
\qquad
\overline{\sigma}
=
\max_{k\le K}\sigma_k.
\]
Thus the theorem gives a local-clock guarantee under arbitrary predictable
allocation, whereas the displayed global-time rate additionally requires
the stated exploration condition.

\paragraph{Comparison with nonasymptotic regret bounds.}
\citet{chen2023recursive} study a different performance criterion for non-private
recursive quantile estimators. For their fixed-budget successive-reject
procedure, Theorem 4.1 therein implies, for fixed $K$, fixed step-size
parameters, and positive suboptimality gaps,
\[
R_n^\tau
:=
\mathbb{E}
\left(
q^\ast-q_{A_{n+1}}
\right)
=
\mathcal{O}
\left(
n^{1+a}
\exp\left(
-c n^{a(1-a)}
\right)
\right),
\]
for some constant $c>0$. This is a non-asymptotic bound
on the expected regret of the final recommendation at a prespecified
sampling budget. In contrast, Theorem \ref{thm:time-uniform-regret} provides an observable bound
that is valid simultaneously over monitoring times, with validity asymptotic
in the deterministic per-arm monitoring start $m$.
The two rates are therefore not directly comparable. The quantity
$R_n^\tau$ measures the expected realized loss of the final recommendation,
whereas $B_t$ measures the remaining uncertainty about the current
recommendation.

The effect of local privacy on the bound is explicit through
\[
\sigma_k^2
=
\frac{\tau(1-\tau)}
{f_k^2(q_k)}
+
\frac{1-r^2}
{4r^2 f_k^2(q_k)}.
\]
The first term is the usual non-private quantile variance, whereas the second
is the additional variance induced by randomized response. The latter
vanishes at $r=1$. Moreover, since
\(
r
=
\tanh\left( {\epsilon_{\mathrm{LDP}}}/{2}\right),
\)
we have $\sigma_k
\sim
\left\{
\epsilon_{\mathrm{LDP}} f_k(q_k)
\right\}^{-1}$ as $\epsilon_{\mathrm{LDP}}\downarrow 0$.
Because this scale is estimated from the privatized trajectories, the
bound does not require knowledge of $f_k(q_k)$ or any additional
privacy budget.

\subsection{Online Quantile A/B Testing}\label{subsec:ab-testing}

A/B tests often target a latency or outcome quantile rather than a mean.
Suppose treatment assignments are predictable from the public transcript and
each assignment reveals a fresh observation from the selected arm. For a
common monitoring start \(m\), construct the delayed-start Gaussian-mixture
interval \(C_{k,n}^{(m)}=[L_{k,n}^{(m)},U_{k,n}^{(m)}]\) at error level
\(\alpha/2\) from the first \(n\) private observations of arm \(k\). At global
time \(t\), let \(N_{k,t}\) be the corresponding local sample size and set
\(\mathcal T_m=\{t:\min_kN_{k,t}\ge m\}\).

The quantile treatment effect is
\(\theta_\tau=Q_2(\tau)-Q_1(\tau)\). For \(t\in\mathcal T_m\), define
\[
C_{t,m}^\Delta
=
\left[
L_{2,N_{2,t}}^{(m)}-U_{1,N_{1,t}}^{(m)},\
U_{2,N_{2,t}}^{(m)}-L_{1,N_{1,t}}^{(m)}
\right].
\]
This is the range of all differences \(y-x\) with
\(x\in C_{1,N_{1,t}}^{(m)}\) and \(y\in C_{2,N_{2,t}}^{(m)}\).

\begin{theorem}[Time-Uniform Quantile-Difference Inference]\label{thm:ab-diff}
Suppose both arms satisfy Assumptions (A1)--(A5). Then
\[
\liminf_{m\to\infty}
\Pp\left(\theta_\tau\in C_{t,m}^\Delta
\text{ for every }t\in\mathcal T_m\right)\ge1-\alpha.
\]
Consequently, under \(H_0:\theta_\tau=\delta_0\),
\[
\limsup_{m\to\infty}
\Pp\left(\exists\,t\in\mathcal T_m:
\delta_0\notin C_{t,m}^\Delta\right)\le\alpha.
\]
\end{theorem}

The sequential test stops at the first \(t\in\mathcal T_m\) for which
\(\delta_0\notin C_{t,m}^\Delta\). Equivalently, after shifting one arm by
\(\delta_0\), the two arm-wise intervals are strictly separated. The
interval-difference construction can be conservative relative to specialized
two-sample martingale tests \citep{howard2022sequential}, but it uses the same
private arm-wise estimator and consumes no additional privacy budget.

\section{Numerical Experiments}\label{sec:experiments}

This section evaluates the finite-sample performance of the
proposed method. Heavy-tailed variance-estimation results, tuning-sensitivity
analyses, and distributional robustness analyses are reported in
Appendix~\ref{app:additional-numerics}.

\subsection{Experimental Design}\label{subsec:simulation-design}

The confidential observations are generated from the standard normal distribution \(N(0,1)\) and the standard Cauchy distribution \(\mathcal C(0,1)\). These two distributions give complementary regimes: the normal case is a regular light-tailed benchmark, whereas the Cauchy case tests robustness to heavy tails while still satisfying the local density assumptions at the target quantiles. The target levels are \(\tau\in\{0.3,0.5,0.8\}\). The truthful-response rates are \(r\in\{1,0.9,0.75,0.5,0.25\}\), corresponding to privacy budgets
\[
\eps=\log(1+r)-\log(1-r)\in\{\infty,2.94,1.95,1.10,0.51\}.
\]
All chains use \(\eta_t=t^{-a}\) with \(a=0.6\); within each repetition,
their common initialization is drawn from \(N(0,1)\). Unless otherwise stated,
the horizon is \(5{,}000{,}000\) and each configuration is repeated \(2000\)
times.
The dynamic number of chains is set to
\[
h(t)=
\begin{cases}
48, & t<1{,}000{,}000,\\
\lfloor 8\log_{10}t\rfloor, & t\ge1{,}000{,}000.
\end{cases}
\]
At every reported time, \(\widehat\sigma_t^2\) is recomputed from all active
chains, including a newly introduced chain during its catch-up phase. We use
the stitched boundary \eqref{eq:stitched-boundary} and the normal-mixture
boundary \eqref{eq:mixture-boundary} with \(\rho=0.001\). The
empirical-quantile confidence sequence of \citet{howard2022sequential} is the
non-private benchmark. We use a short burn-in of approximately
\((0.25/r^2)\%\) of the horizon.

\subsection{Performance of the Inferential Procedure}
\label{subsec:simulation-performance}

\begin{figure}[htbp]
\centering
\includegraphics[width=0.90\textwidth]{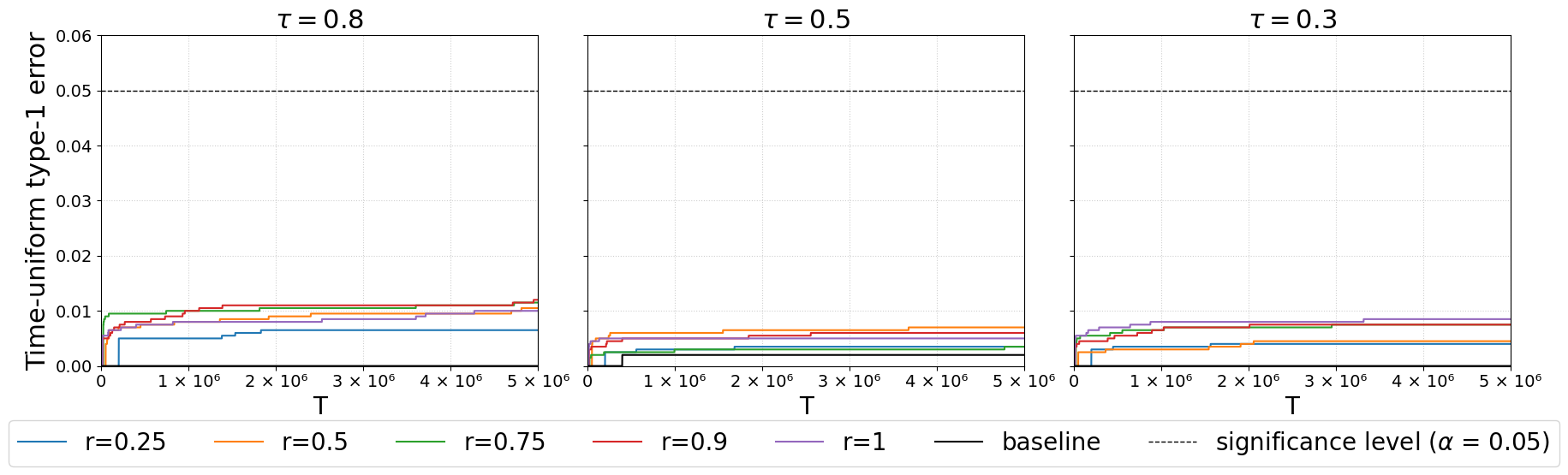}
\includegraphics[width=0.90\textwidth]{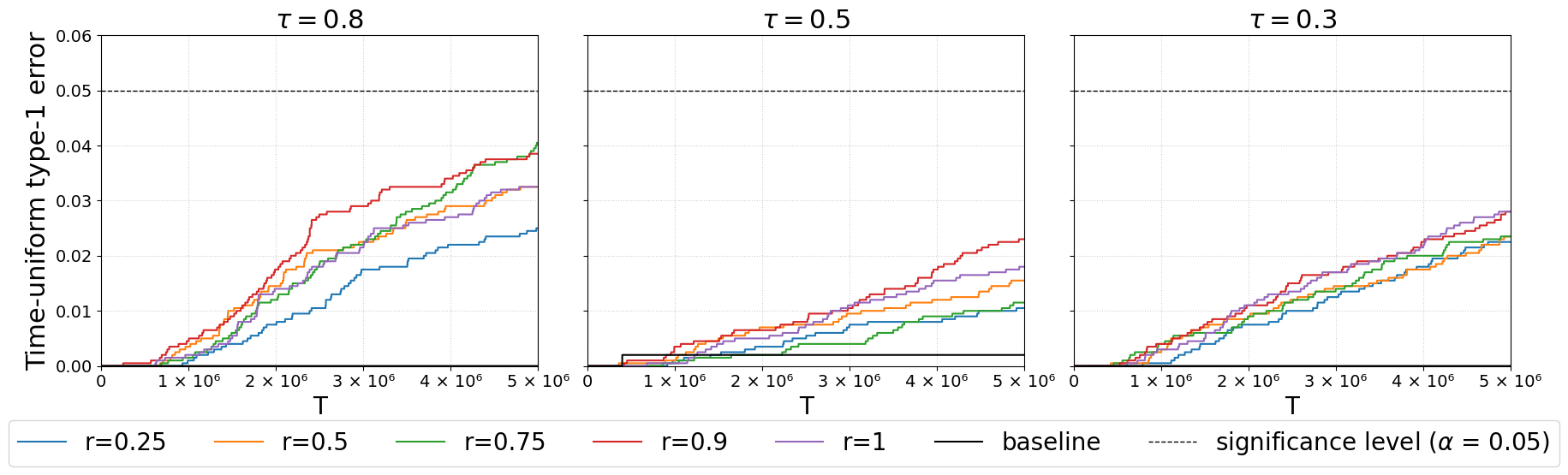}
\caption{Empirical anytime type-I error for 95\% AsympCSs under \(N(0,1)\). Top: stitched boundary \eqref{eq:stitched-boundary}. Bottom: mixture boundary \eqref{eq:mixture-boundary}. The non-private benchmark is from \citet{howard2022sequential}.}
\label{fig:type1-normal}
\end{figure}

\begin{figure}[htbp]
\centering
\includegraphics[width=0.90\textwidth]{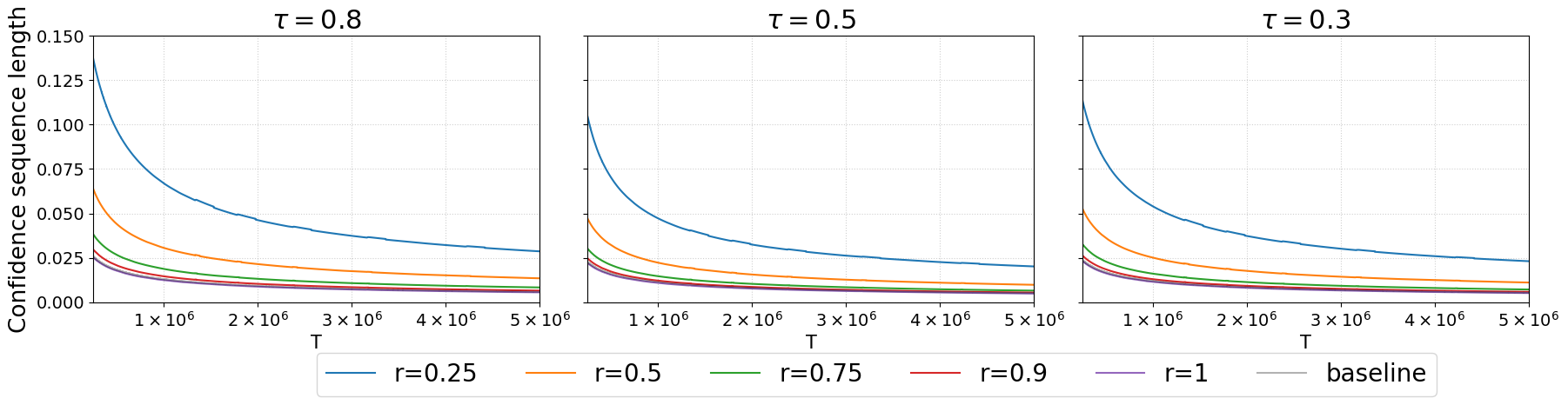}
\includegraphics[width=0.90\textwidth]{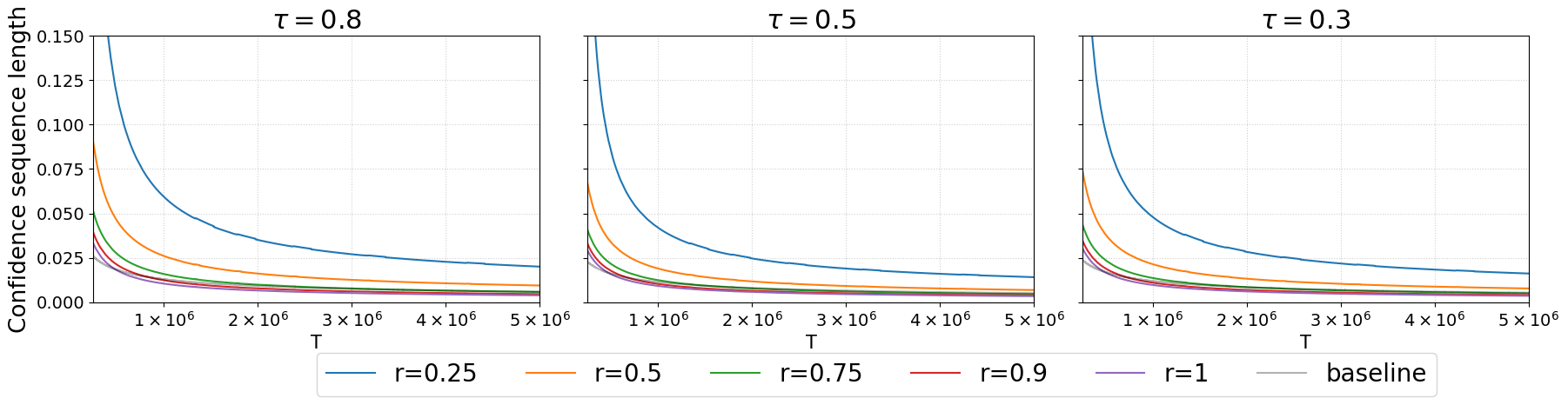}
\caption{Average confidence-sequence width for the same normal-distribution settings as Figure~\ref{fig:type1-normal}. Top: stitched boundary. Bottom: mixture boundary.}
\label{fig:width-normal}
\end{figure}

Under this design, we evaluate the inferential procedure
along three dimensions: anytime coverage and confidence-sequence width,
variance-estimation accuracy, and fixed-time efficiency.
Figures~\ref{fig:type1-normal} and~\ref{fig:width-normal} report the
empirical anytime error and average confidence-sequence width
under the normal distribution.

At each monitored time, the empirical
anytime type-I error is the proportion of repetitions in which the true
quantile has left the reported confidence sequence at least once up to that
time, and the average width is computed across repetitions.
The empirical anytime error remains close to or below
the nominal \(5\%\) level across the reported settings. Confidence-sequence width
increases as \(r\) decreases, in agreement with \eqref{eq:sigma2}. The stitched
boundary has the sharper asymptotic LIL order, whereas the mixture
boundary is generally shorter over the simulated horizon. At \(r=1\), the
stochastic-approximation AsympCSs are competitive with the exact non-private
empirical-quantile benchmark.

\begin{figure}[htbp]
\centering
\includegraphics[width=0.80\textwidth]{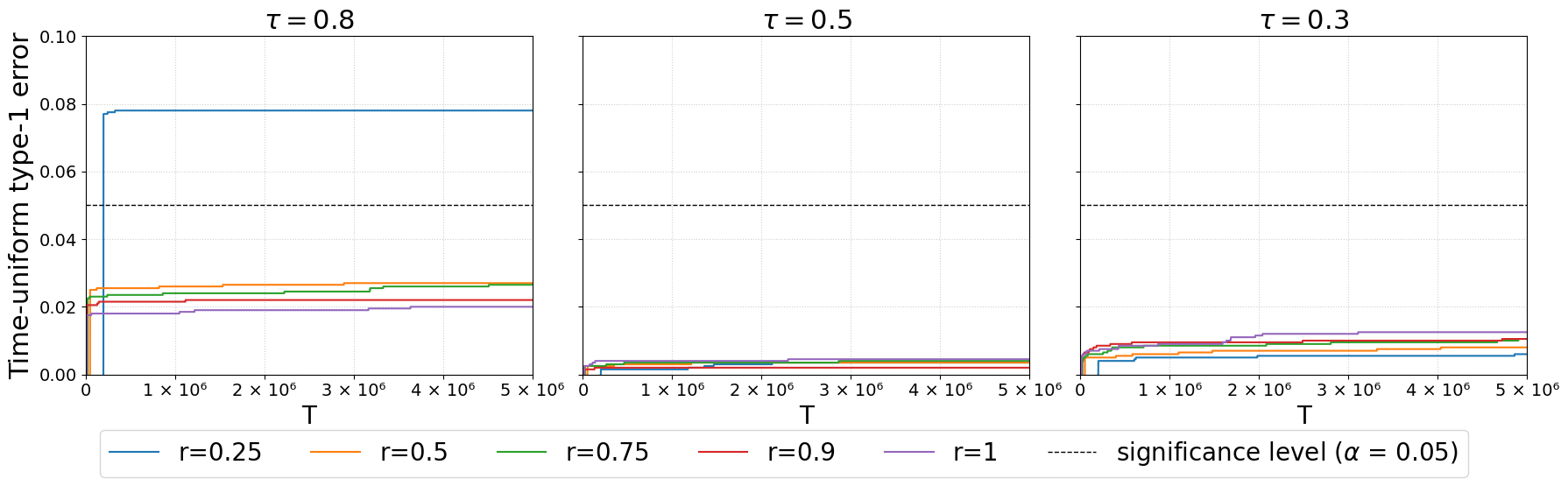}
\includegraphics[width=0.80\textwidth]{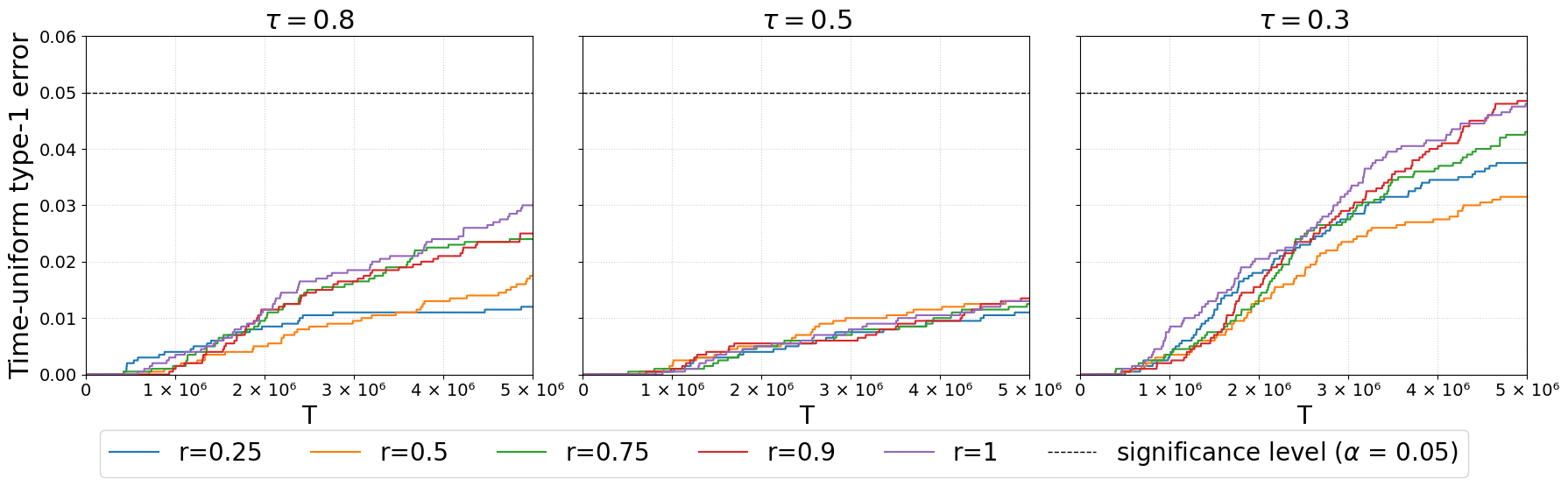}
\caption{Empirical anytime type-I error for 95\% AsympCSs under the standard Cauchy distribution. Top: stitched boundary. Bottom: mixture boundary.}
\label{fig:type1-cauchy}
\end{figure}

\begin{figure}[htbp]
\centering
\includegraphics[width=0.80\textwidth]{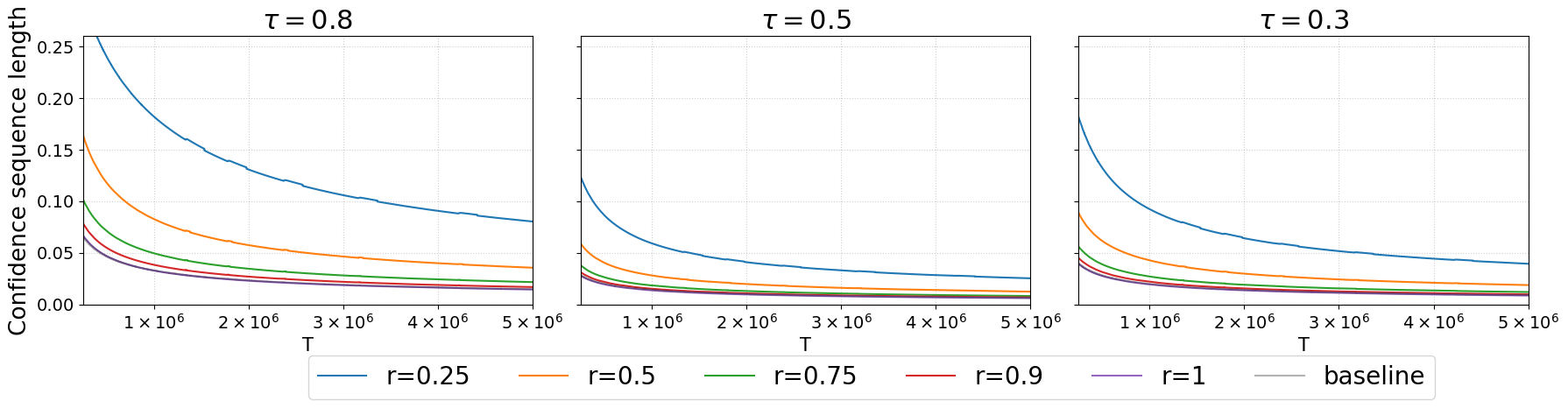}
\includegraphics[width=0.80\textwidth]{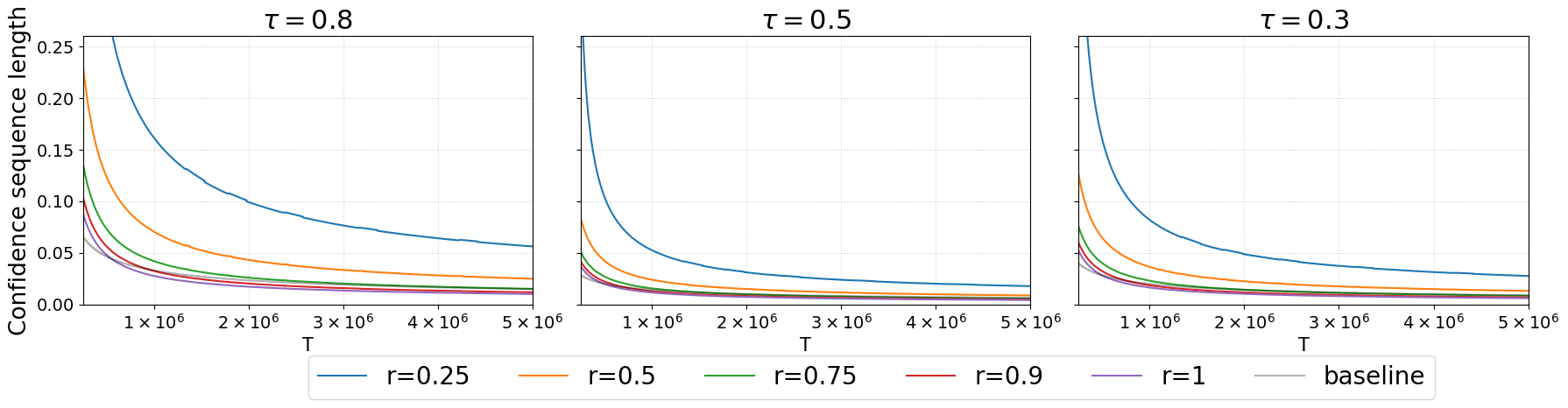}
\caption{Average confidence-sequence width under the standard Cauchy distribution. Top: stitched boundary. Bottom: mixture boundary.}
\label{fig:width-cauchy}
\end{figure}

For the Cauchy design, the same privacy--width tradeoff holds
(Figures~\ref{fig:type1-cauchy} and~\ref{fig:width-cauchy}), and empirical
anytime error is at or below the nominal \(5\%\) level in almost all reported
configurations.

\begin{figure}[htbp]
\centering
\includegraphics[width=0.90\textwidth]{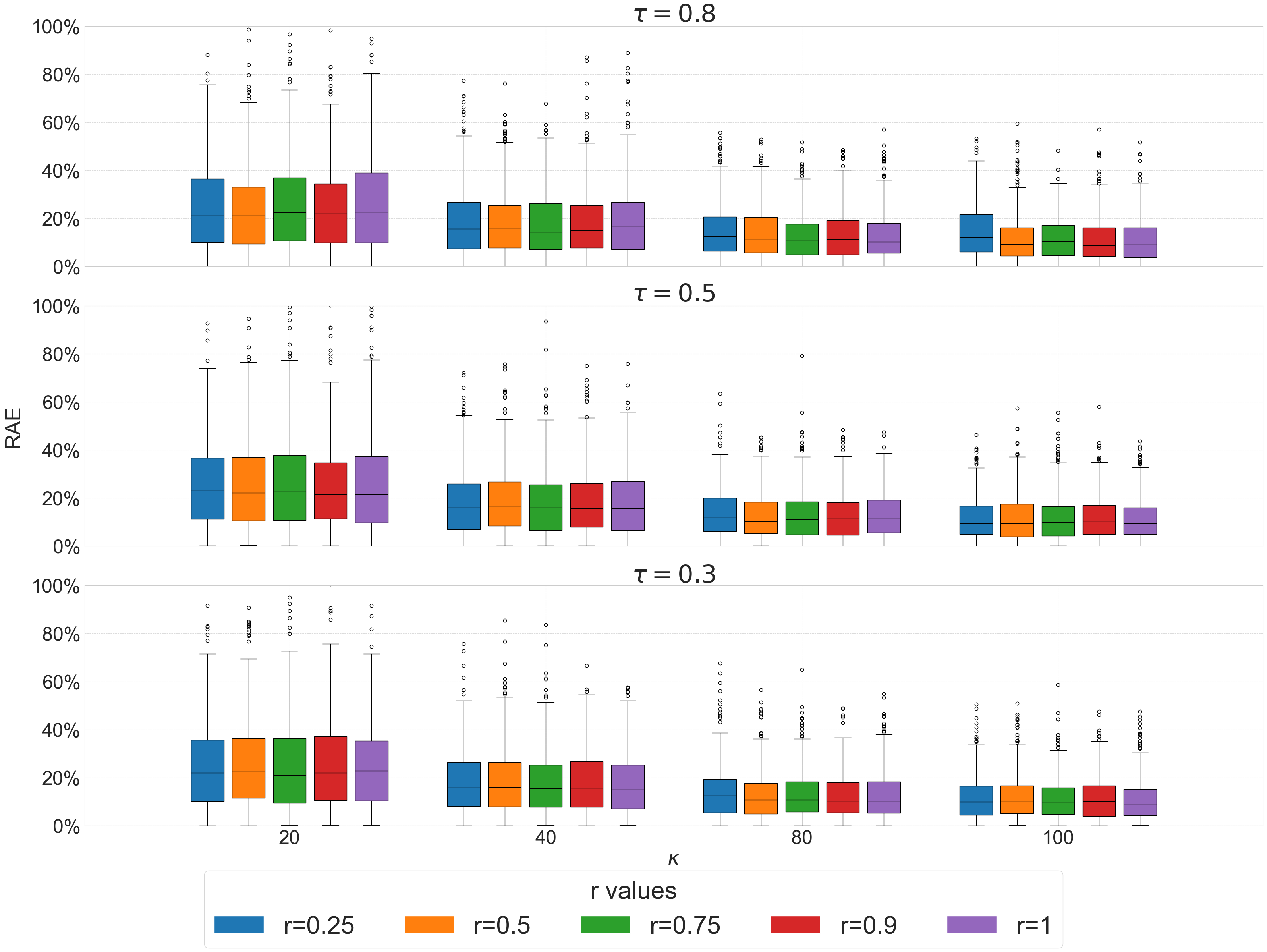}
\caption{Relative absolute error of \(\widehat\sigma_T^2\) under the standard normal distribution for fixed chain counts \(\kappa\in\{20,40,80,100\}\).}
\label{fig:rae-normal}
\end{figure}

We next evaluate the finite-sample
accuracy of the plug-in variance estimator \eqref{eq:variance-estimator}. We
hold the number of chains fixed at \(\kappa=20,40,80\), or \(100\), and compute
the relative absolute error
\[
\operatorname{RAE}=\frac{|\widehat\sigma_T^2-\sigma^2|}{\sigma^2}.
\]
Varying fixed \(\kappa\) isolates the effect of cross-chain replication, which changes only a few times under the dynamic schedule over this horizon. Figure~\ref{fig:rae-normal} shows that the RAE is comparable across privacy levels at a given chain count and decreases as more chains are used, in agreement with the cross-chain averaging term in \eqref{eq:variance-general-rate}. The corresponding Cauchy RAE results are reported in Appendix~\ref{app:cauchy-variance}.

\begin{figure}[htbp]
\centering
\includegraphics[width=0.90\textwidth]{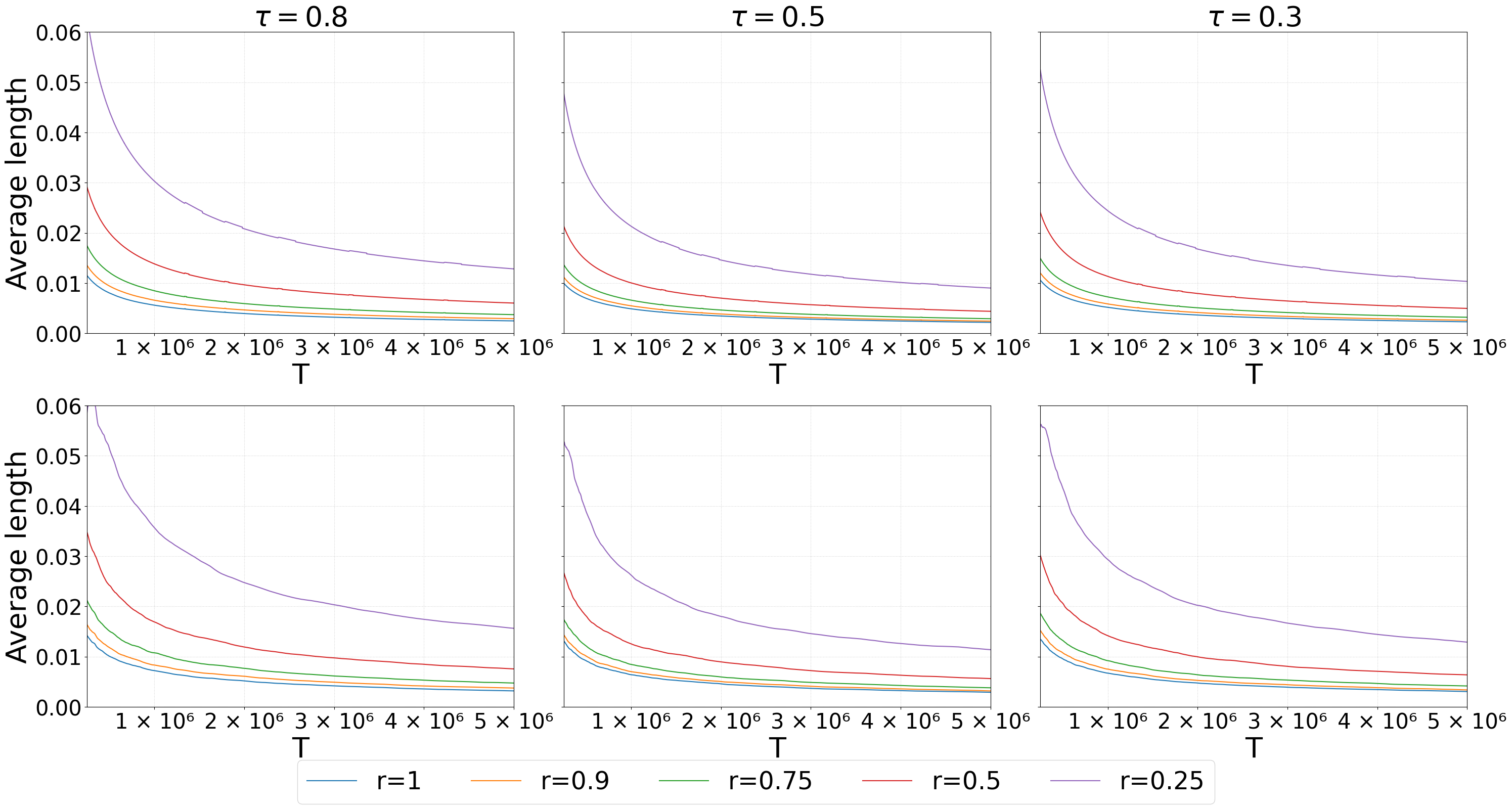}
\caption{Average pointwise interval length for Corollary~\ref{cor:pointwise-ci} and for the self-normalized interval of \citet{liu2023online}.}
\label{fig:sn-length}
\end{figure}

\begin{figure}[htbp]
\centering
\includegraphics[width=0.90\textwidth]{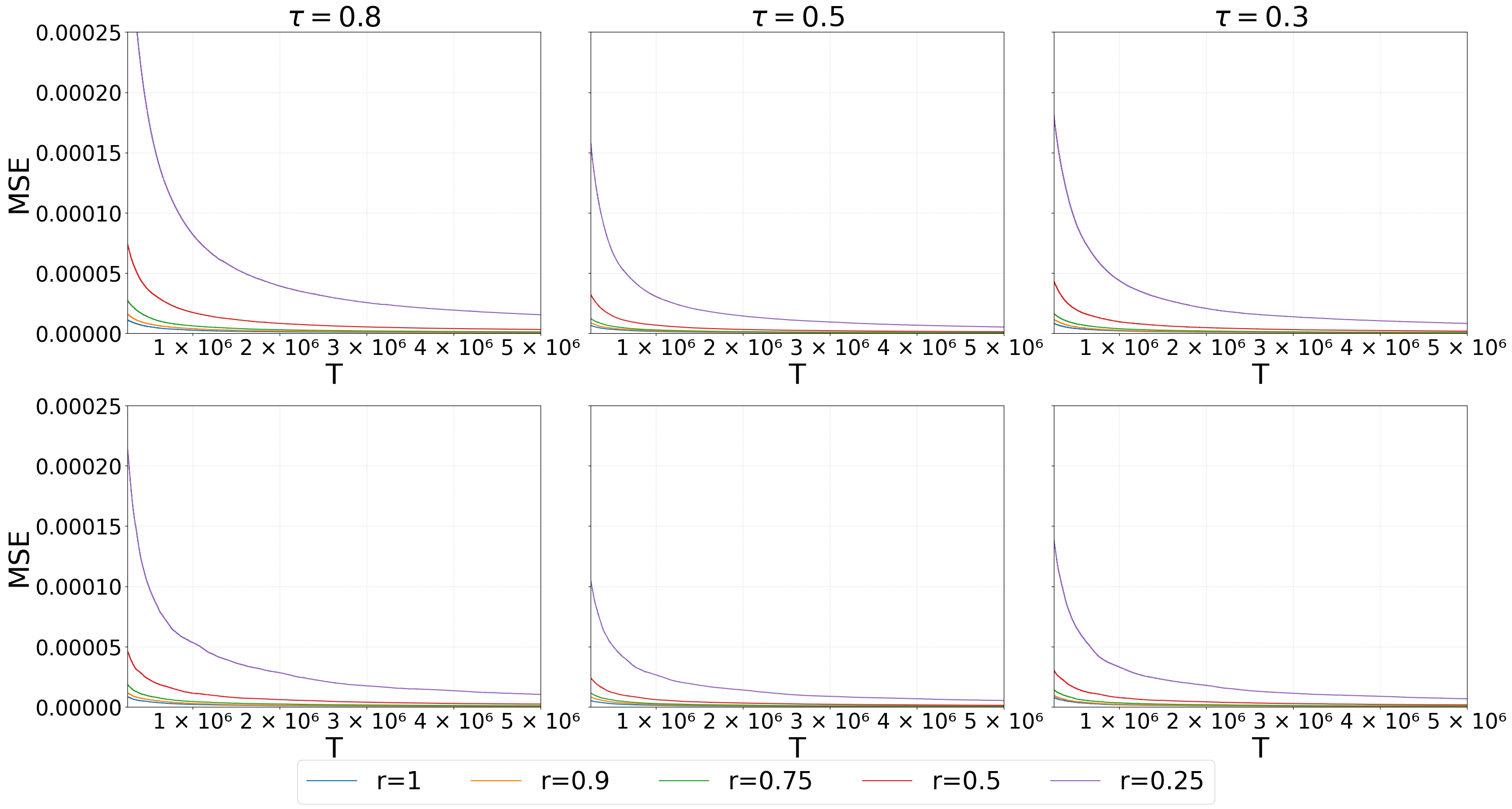}
\caption{Mean squared error comparison for point estimation under the same normal-distribution settings.}
\label{fig:mse-comparison}
\end{figure}

To assess fixed-time efficiency, we compare the fixed-time interval in
Corollary~\ref{cor:pointwise-ci} with the self-normalized LDP interval of
\citet{liu2023online} under the same normal settings. Figures~\ref{fig:sn-length} and~\ref{fig:mse-comparison} show that both the average interval length and the MSE decrease as \(T\) increases and increase as \(r\) decreases. Additional tuning-sensitivity and distributional-robustness results are
reported in Appendix~\ref{app:sensitivity-robustness}.

\subsection{Sequential Decision Experiments}
\label{subsec:sequential-experiments}

The sequential-decision experiments evaluate the two
procedures developed in Section~\ref{sec:applications}, beginning with
quantile best-arm identification. We vary the target quantile over the grid
$
\tau \in \{0.3,0.4,0.5,0.6,0.7\}.
$
We consider $K=4$ arms. One exceptional arm has distribution $N(0,2^2)$, and the
remaining $K-1$ arms have distribution $N(0,1)$.
Following \citet{howard2022sequential}, we use a quantile-level perturbation
at $0.025$, that is, we consider the interval $(\tau-0.025,\tau+0.025)$.
The proposed stopping rule continues to compare AsympCSs for
\(Q_k(\tau)\), in accordance with the retained definition of
\(\epsilon\)-optimality.  We set
\[
\epsilon(\tau)
=
\Bigl(
Q_{k^\star(\tau)}(\tau)-Q_{k^\star(\tau)}(\tau-0.025)
+
\max_{j\neq k^\star(\tau)}
\{Q_j(\tau+0.025)-Q_j(\tau)\}
\Bigr),
\]
where
$
k^\star(\tau)\in\arg\max_{1\le k\le K} Q_k(\tau).
$
The number of chains for each arm follows the plateau-adjusted version of
$
h(n)=4+\lfloor 7\bigl(n^{0.1}-1\bigr)\rfloor,
$ where \(n\) is the arm-local effective sample size; the resulting
unit-jump schedule \(h(n)\asymp n^{0.1}\) satisfies
\eqref{eq:plateau-schedule} and Assumptions~(A4)--(A5).
For our method, the confidence-sequence bounds $L^\tau_{k,t}$ and $U^\tau_{k,t}$ are
constructed using the   Gaussian-mixture boundary in
\eqref{eq:delayed-mixture-boundary}, with each complete two-sided arm-wise AsympCS calibrated at error level \(\delta/K=0.0125\).
For each arm, we perform a burn-in stage before constructing the AsympCSs.
In addition, adaptive LUCB comparisons begin only after every arm has accumulated
$m=1500$ effective post-burn-in observations to improve algorithmic stability.
The numbers of discarded per-chain burn-in iterations are $180$, $200$, $350$, and $1200$ for
$r=1.0,0.8,0.6,0.4$, respectively.
For comparison, we use the non-private Howard--Ramdas quantile-level QLUCB
comparator, which operates directly on empirical quantiles. The two stopping
criteria use quantile-level and quantile-value tolerances, respectively, with
the above choice of \(\epsilon(\tau)\) relating their scales in this simulation.
The maximum sampling budget is $500{,}000$ pulls, and each configuration is replicated
$1000$ times.

We report three quantities.
The first is the average sample size, defined as the mean total number of
arm pulls until stopping.
The second is the
empirical \(\epsilon\)-optimal selection rate, defined as the
proportion of repetitions in which the selected arm is
\(\epsilon\)-optimal.
The third is the joint pre-stopping miscoverage rate, defined as the
proportion of runs in which at least one true arm quantile leaves its
arm-wise AsympCS before the algorithm stops.

\begin{figure}[htbp]
\centering
\includegraphics[width=\textwidth]{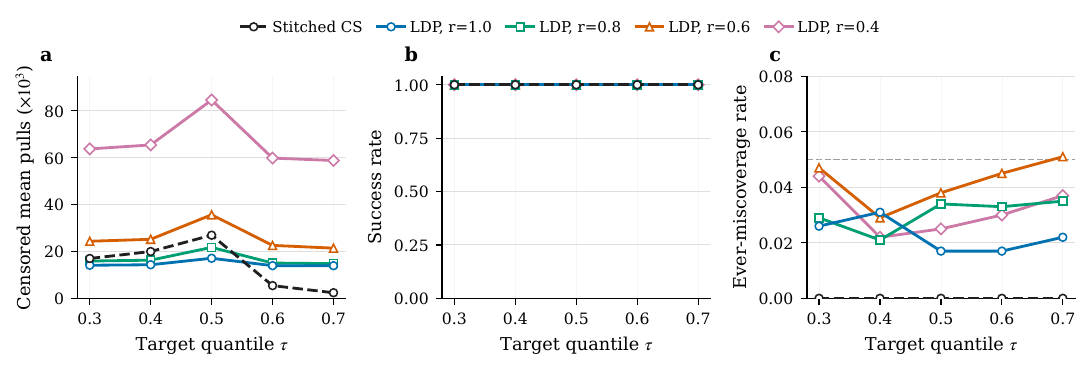}
\caption{Quantile \(\epsilon\)-best-arm identification in the normal-spread
setting, with one arm following \(N(0,2^2)\), the remaining three arms
following \(N(0,1)\), and
\(\tau\in\{0.3,0.4,0.5,0.6,0.7\}\). Panel a reports the average number of
arm pulls until stopping. Panel b reports the
empirical \(\epsilon\)-optimal selection rate, and Panel c
reports the joint pre-stopping miscoverage rate.  The dashed black curve
denotes the non-private Howard--Ramdas quantile-level QLUCB comparator.}
\label{fig:ldp_best_arm_normal_panel}
\end{figure}

Figure~\ref{fig:ldp_best_arm_normal_panel}
reports the three operating characteristics across the
quantile levels and truthful-response rates considered.
First, for every reported configuration, all \(1000\)
repetitions stop before reaching the maximum sampling budget.
The empirical \(\epsilon\)-optimal selection rate is \(100\%\)
in each reported configuration, while the
joint pre-stopping
miscoverage rates do not exceed \(5.0\%\) in almost all configurations. These
finite-sample findings are
qualitatively aligned with the asymptotic time-uniform validity established in
Theorem~\ref{thm:QLUCB}.
Second, the average number of arm pulls is largest near \(\tau=0.5\), where
the \(\epsilon\)-optimal set changes and the effective gaps are smallest.
At \(\tau=0.6\) and \(\tau=0.7\), the exceptional arm is more clearly
separated, so the LUCB rule stops substantially earlier. The number of pulls
also generally increases as the truthful-response rate decreases. These
patterns are consistent with the interval-separation mechanism in the proof
of Theorem~\ref{thm:QLUCB}: smaller reward gaps and larger privacy-inflated arm
scales make the stopping condition harder to satisfy.
Third, the non-private comparator requires the fewest samples when
\(\tau\ge0.6\), where the problem is relatively easy and the initialization
cost of the LDP procedure is more prominent. For \(\tau\le0.5\), however,
the proposed method with \(r=1.0\) requires fewer samples than the comparator,
and a similar advantage is observed for \(r=0.8\) over most of this range.

The best-arm experiment uses adaptive allocation across four streams. The online quantile A/B experiment instead uses deterministic
alternating allocation. We fix the target
quantile level and vary the true quantile treatment effect. Specifically,
\[
X_1\sim N(0,1),\qquad X_2\sim N(\Delta,1),
\]
so that \(Q_2(\tau)-Q_1(\tau)=\Delta\) for every \(\tau\).  We set \(\tau=0.5\),
\[
\Delta\in\{0,0.02,0.04,0.06,0.08,0.10,0.15,0.20,0.30\},
\]
\(r\in\{1.0,0.8,0.6,0.4\}\), and \(\alpha=0.05\). We set
\(T_{\max}=100{,}000\) total arm pulls across the two arms. The non-private comparator is the
Howard--Ramdas stitched confidence-sequence baseline: it constructs empirical arm-wise
stitched quantile confidence sequences and then applies the same interval-difference
stopping rule.
Each arm uses the same plateau-adjusted polynomial dynamic-chain schedule as
in the preceding best-arm experiment. The complete two-sided arm-wise AsympCSs are calibrated at error level \(\alpha/2=0.025\). For the LDP
procedures, monitoring begins after each arm has accumulated \(m=1500\)
effective post-burn-in observations. The two arms are then sampled in strict
alternation. Every chain
discards its first \(180,200,350,1200\) updates for
\(r=1.0,0.8,0.6,0.4\), respectively; these discarded reports are included in
the stopping cost. We report the probability of rejection by the horizon
under the null and alternatives, together with the mean stopping time censored
at the horizon.

\begin{figure}[htbp]
  \centering
  \includegraphics[width=\textwidth]{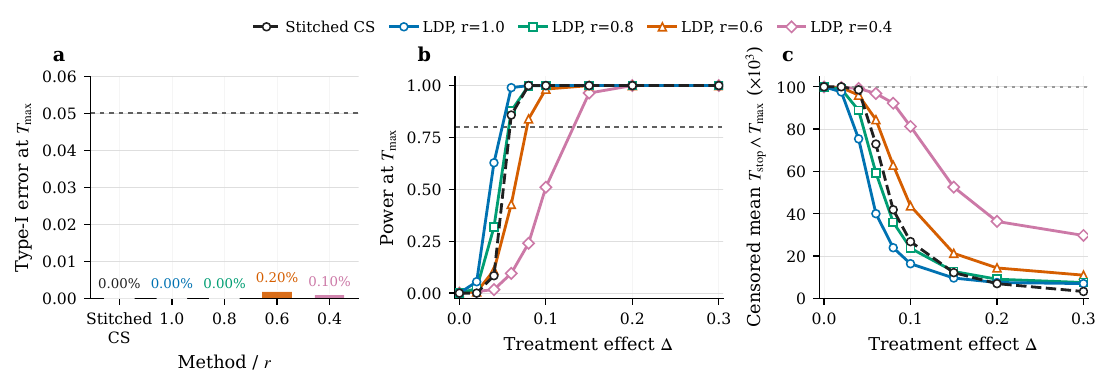}
\caption{Online quantile A/B testing in the normal-shift setting. Panel a
reports the empirical probability of ever rejecting by \(T_{\max}\) under
the exact null \(\Delta=0\). Panel b reports the finite-horizon empirical
power as a function of the quantile treatment effect \(\Delta\). Panel c
reports the censored mean stopping time
\(T_{\rm stop}\wedge T_{\max}\). The LDP
procedures use monitoring start \(m=1500\), and each method--effect
configuration is evaluated over \(1000\)
independent repetitions. The black series denotes the exact non-private
Howard--Ramdas stitched confidence-sequence baseline.}
  \label{fig:ab-power}
\end{figure}

Figure~\ref{fig:ab-power} summarizes the results based on \(1000\)
independent repetitions for each method--effect configuration.
First, under the exact null, the empirical probability of ever rejecting by
\(T_{\max}\) ranges from \(0\%\) to \(0.20\%\), substantially below the
nominal level for all methods. This conservative null behavior reflects the
interval-difference construction and is consistent with the asymptotic time-uniform
arm-wise coverage in Corollary~\ref{cor:gm-sharp-coverage} together with
Theorem~\ref{thm:ab-diff}.
Second, under the alternatives, power increases with the treatment effect
\(\Delta\) and decreases as the truthful-response rate \(r\) decreases. The
empirical \(80\%\)-power thresholds are \(0.06\) for the non-private
baseline and for \(r=1.0\) and \(r=0.8\), \(0.08\) for \(r=0.6\), and
\(0.15\) for \(r=0.4\). No wrong-direction rejections are observed over the
reported alternative grid.
Third, the censored mean stopping time decreases with \(\Delta\) and
increases under stronger privacy.
Taken together, the reported operating characteristics
quantify the finite-horizon tradeoff between stronger local privacy and the
empirical power and censored mean stopping time of the arm-wise LDP
quantile-difference procedure.

\section{Salary Data Analysis}\label{sec:real-data}

We analyze annual salaries from the government-salary data studied by
\citet{plevcko2024fairadapt}, which are derived from the 2018 American Community
Survey. Salary is positive, skewed, and sensitive. The offline study emulates
the local protocol: each record is used to generate one private response, and
the analyst-side recursion thereafter uses only those responses.  A second illustration using
law-school GPA is included in Appendix~\ref{app:law-school}.

\begin{figure}[htbp]
\centering
\includegraphics[width=0.90\textwidth]{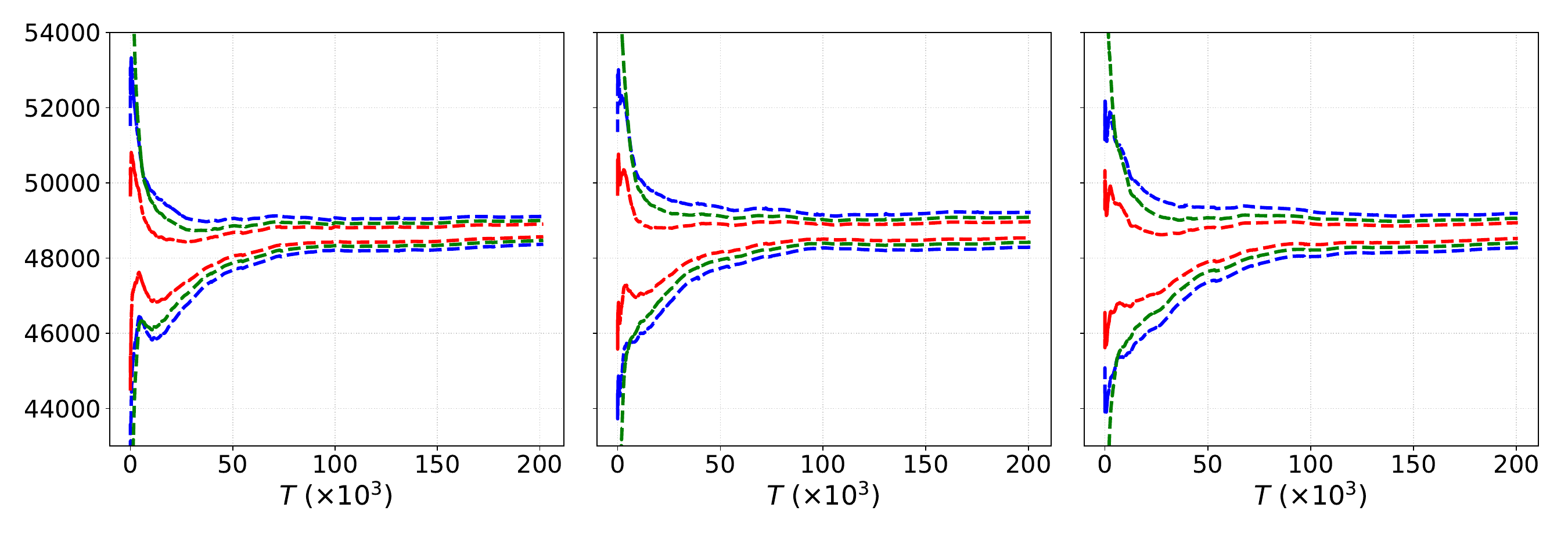}
\caption{Pointwise median confidence intervals and median AsympCSs for annual salary in the government salary data. Each panel compares the pointwise interval from Corollary~\ref{cor:pointwise-ci}, the stitched AsympCS and the mixture AsympCS.}
\label{fig:real-data}
\end{figure}

For annual salary \(S\) in dollars, we apply the recursion to \(Y=\log(1+S/1000)-4\) and back-transform the displayed bounds using \(S=1000\{\exp(Y+4)-1\}\).
The target is the median, \(\tau=0.5\), and
\(r\in\{0.9,0.8,0.75\}\). The mixture boundary uses \(\rho=0.01\); the
remaining choices follow Section~\ref{subsec:simulation-design}.
Figure~\ref{fig:real-data}
shows that stronger privacy produces wider AsympCSs, as predicted by
\eqref{eq:sigma2}. Over this horizon the mixture boundary is shorter, while
the stitched boundary retains the sharper asymptotic order.

We next use the same data for two sequential decisions. Both analyses use the
arm-local dynamic-chain implementation and  Gaussian-mixture
AsympCSs from Section~\ref{sec:applications}.

\begin{figure}[htbp]
\centering
\includegraphics[width=0.86\textwidth]{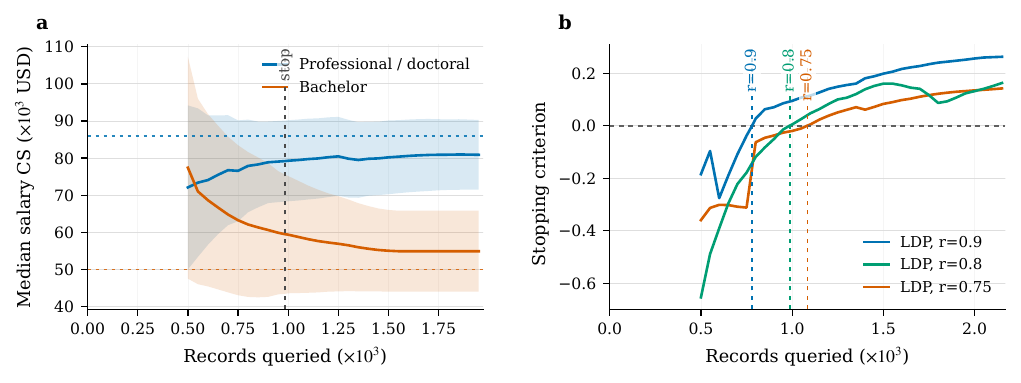}
\caption{Representative salary case study for LDP quantile
\(\epsilon\)-best-arm identification. Panel a shows the back-transformed
median-salary AsympCSs for the selected education group and its strongest
challenger at \(r=0.8\); horizontal dashed lines mark
full-data empirical medians used only for interpretation. Panel b shows the
stopping margin for \(r=0.9,0.8,0.75\); the horizontal dashed line marks zero, and, for each \(r\),
the first upward crossing triggers the LUCB stopping rule. Vertical dashed
lines mark the corresponding stopping times. Trajectories after stopping are
displayed only to visualize subsequent AsympCS shrinkage.}
\label{fig:salary-bestarm-case}
\end{figure}

We first apply Quantile-LUCB to these salary data, using the transformation defined above.
The algorithm is run on the transformed scale, while the arm-wise AsympCSs in Figure~\ref{fig:salary-bestarm-case} are back-transformed to dollars for interpretation. The arms are five pre-specified education groups: high school or less, some college/associate, bachelor, master, and professional/doctoral.
We set \(\tau=0.5\), \(\epsilon=0.10\) on the transformed scale,
\(\delta=0.05\), and a monitoring start of \(m=100\) effective observations
per arm. Each arm uses the plateau-adjusted dynamic-chain schedule, zero
burn-in, and a  Gaussian-mixture AsympCS at arm-wise error
level \(\delta/K=0.01\). We consider \(r\in\{0.9,0.8,0.75\}\).
Full-data empirical medians are used as descriptive references.

Figure~\ref{fig:salary-bestarm-case} illustrates how the proposed arm-wise
LDP AsympCSs can be converted into an actionable sequential
selection rule. The corresponding representative arm-wise trajectories for all
three truthful-response rates are reported separately in
Appendix~\ref{app:salary-trajectories}.
In all three representative runs, Quantile-LUCB selects the
professional/doctoral group, which also has the highest descriptive full-data
empirical median, and stops after \(728\), \(994\), and \(1084\) queried
records for \(r=0.9,0.8,0.75\), respectively. In addition, the algorithm stops in every run
and selects this group in \(90.5\%\), \(88.0\%\), and \(84.5\%\) of the runs,
respectively. The ordering agrees with the descriptive full-data medians.
Stopping takes longer and selection is less stable as \(r\) decreases,
reproducing the privacy--separation tradeoff seen in the simulation.

We also apply the online quantile A/B test to the same government salary data. We test equality of regional median salaries,
\[
H_0:Q_{\rm Mideast}(0.5)-Q_{\rm Southeast}(0.5)=0,
\]
where Southeast is treated as the control arm and Mideast as the treatment arm. We use the same log-salary transformation as above, run the LDP AsympCSs on the transformed scale, and back-transform the displayed arm-wise bounds to dollars.
We set \(\alpha=0.05\) and use a monitoring start of \(m=100\)
effective observations per arm. Each arm uses the plateau-adjusted
dynamic-chain schedule, zero burn-in, and a  Gaussian-mixture
AsympCS at arm-wise error level \(\alpha/2=0.025\). We consider
\(r\in\{0.9,0.8,0.75\}\).

\begin{figure}[htbp]
\centering
\includegraphics[width=0.86\textwidth]{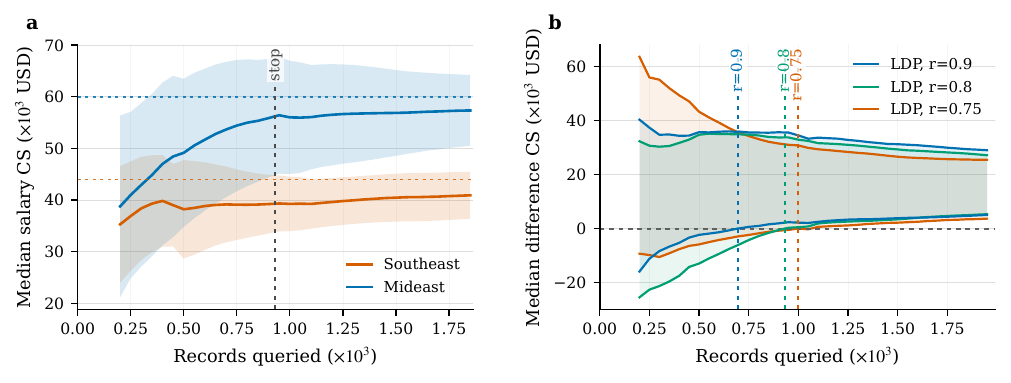}
\caption{Representative salary case study for LDP online
quantile A/B testing. Panel a shows back-transformed arm-wise median-salary
AsympCSs for Southeast and Mideast at \(r=0.8\); horizontal dashed lines mark the corresponding full-data
empirical medians used only for interpretation. Panel b shows the induced dollar-scale delayed-start family for
\(Q_{\rm Mideast}(0.5)-Q_{\rm Southeast}(0.5)\) under
\(r=0.9,0.8,0.75\). The horizontal dashed line marks zero,
and vertical dashed lines mark the first times at which zero is excluded.
Trajectories after stopping are displayed only to visualize subsequent interval-width
shrinkage.}
\label{fig:salary-ab-case}
\end{figure}

Figure~\ref{fig:salary-ab-case}, together with the representative arm-wise
trajectories for all three truthful-response rates in
Appendix~\ref{app:salary-trajectories}, illustrates the
sequential comparison of regional median salaries. In each representative
run, the induced median-difference family excludes zero and
indicates a higher median salary in the Mideast region, with stopping times of
\(696\), \(932\), and \(978\) queried records for
\(r=0.9,0.8,0.75\), respectively. In addition, the procedure rejects the equal-median null by the
analysis horizon in every run and concludes that the Mideast median is higher
in \(97.0\%\), \(95.5\%\), and \(95.5\%\) of the runs, respectively. The
directional conclusion is stable across the reported privacy levels, while the
later stopping and slightly lower agreement at smaller \(r\) quantify the
cost of stronger local randomization.

\section{Concluding Remarks}\label{sec:discussion}

Using dynamically expanding parallel SGD chains, we construct an online LDP quantile estimator together with a variance estimator based solely on privatized iterates. We establish a strong Gaussian approximation and strong consistency of the variance estimator, which yield asymptotic confidence sequences and  asymptotic time-uniform coverage. These results lead to sequential procedures for quantile best-arm identification and A/B testing. The simulations and salary-data applications illustrate the finite-sample privacy--accuracy tradeoff.

Two extensions remain open. First, finite-sample calibration would require explicit time-uniform bounds for the nonsmooth stochastic-approximation remainder, chain-entry transients, and scale-estimation error. Second, simultaneous inference over quantile levels would require a joint private encoding, a uniform-in-level approximation, and consistent estimation of cross-quantile covariance.


\appendix

\section{Additional Numerical Experiments}
\label{app:additional-numerics}

\subsection{Heavy-Tailed Variance-Estimation Accuracy}
\label{app:cauchy-variance}

\begin{figure}[htbp]
\centering
\includegraphics[width=0.90\textwidth]{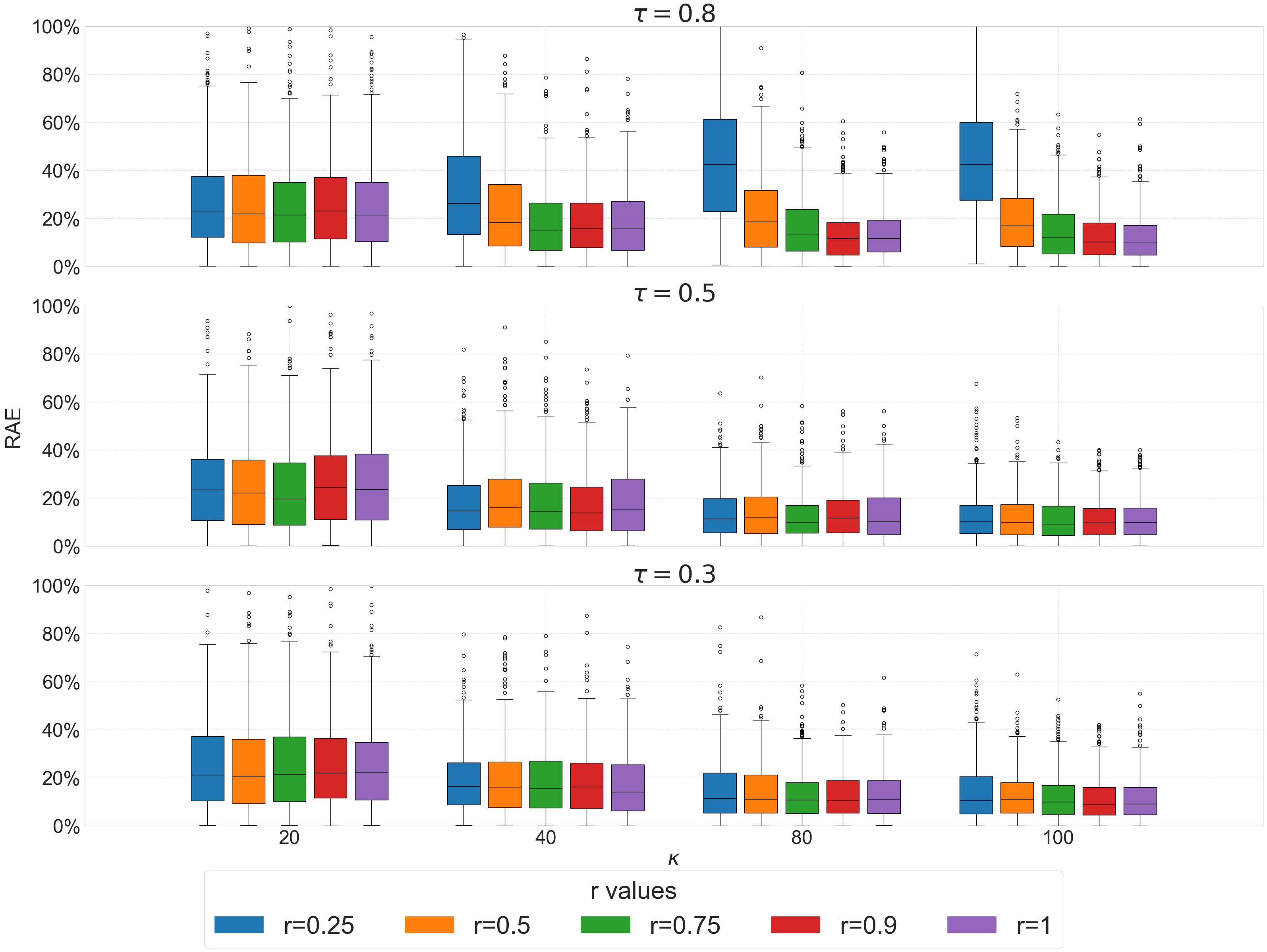}
\caption{Relative absolute error of \(\widehat\sigma_T^2\) under the standard Cauchy distribution for fixed chain counts \(\kappa\in\{20,40,80,100\}\).}
\label{fig:rae-cauchy}
\end{figure}

The corresponding RAE results under the standard Cauchy distribution are reported in Figure~\ref{fig:rae-cauchy}.

\subsection{Sensitivity and Distributional Robustness}
\label{app:sensitivity-robustness}

The method has three types of tuning parameters. The first type is inherited from SGD, such as the step-size exponent \(a\). The second type is inherited from time-uniform inference, such as the start time \(m\) and the mixture-boundary parameter \(\rho\). The third type is specific to the proposed method, such as the chain-growth function \(h(t)\). We study all three types under \(T=5{,}000{,}000\), \(r=0.75\), \(\tau=0.5\), normal data and 1000 repetitions.

\begin{figure}[htbp]
\centering
\includegraphics[width=0.90\textwidth]{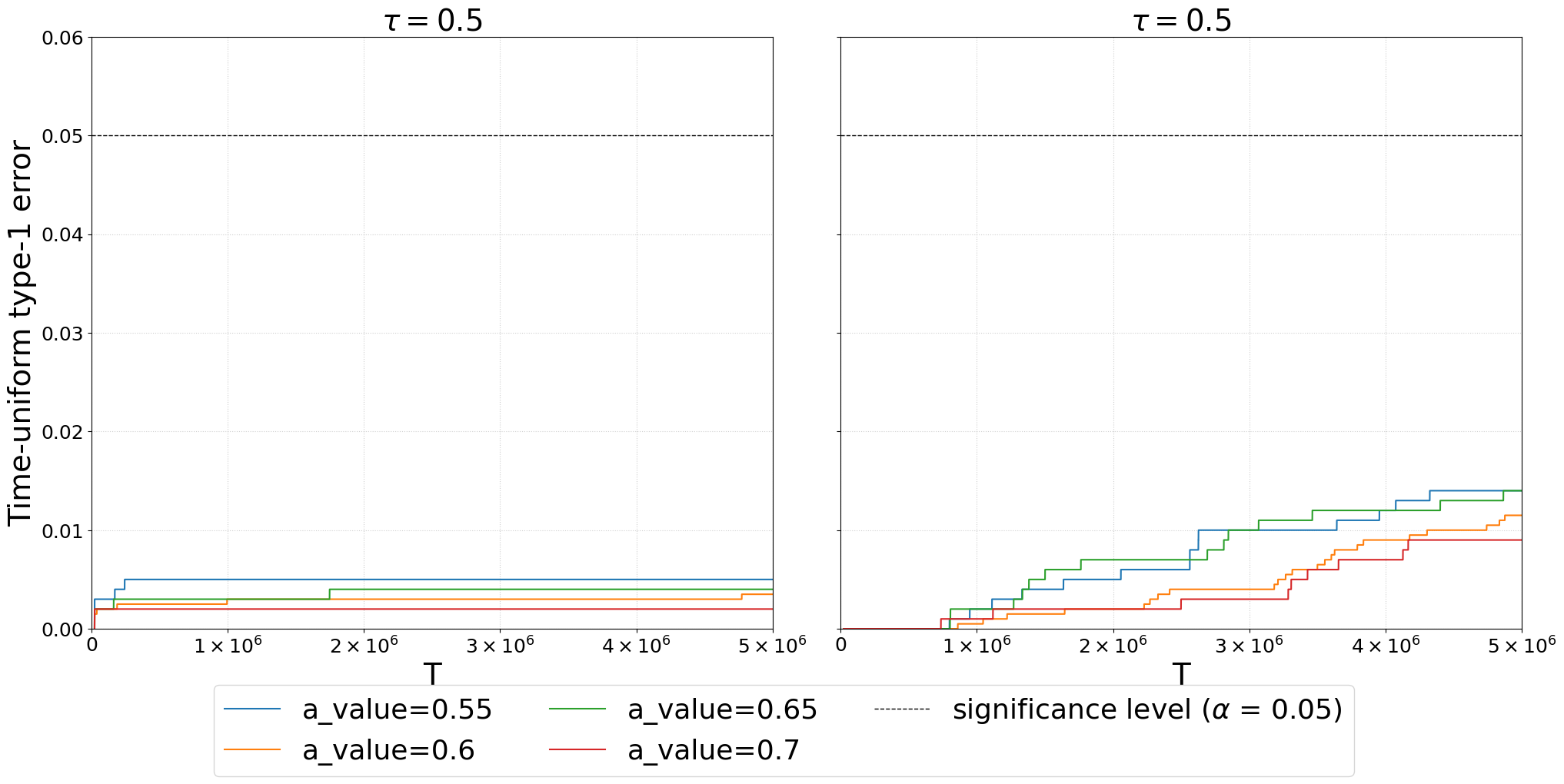}
\caption{Sensitivity to the step-size exponent \(a\).}
\label{fig:sens-a}
\end{figure}

\begin{figure}[htbp]
\centering
\includegraphics[width=0.90\textwidth]{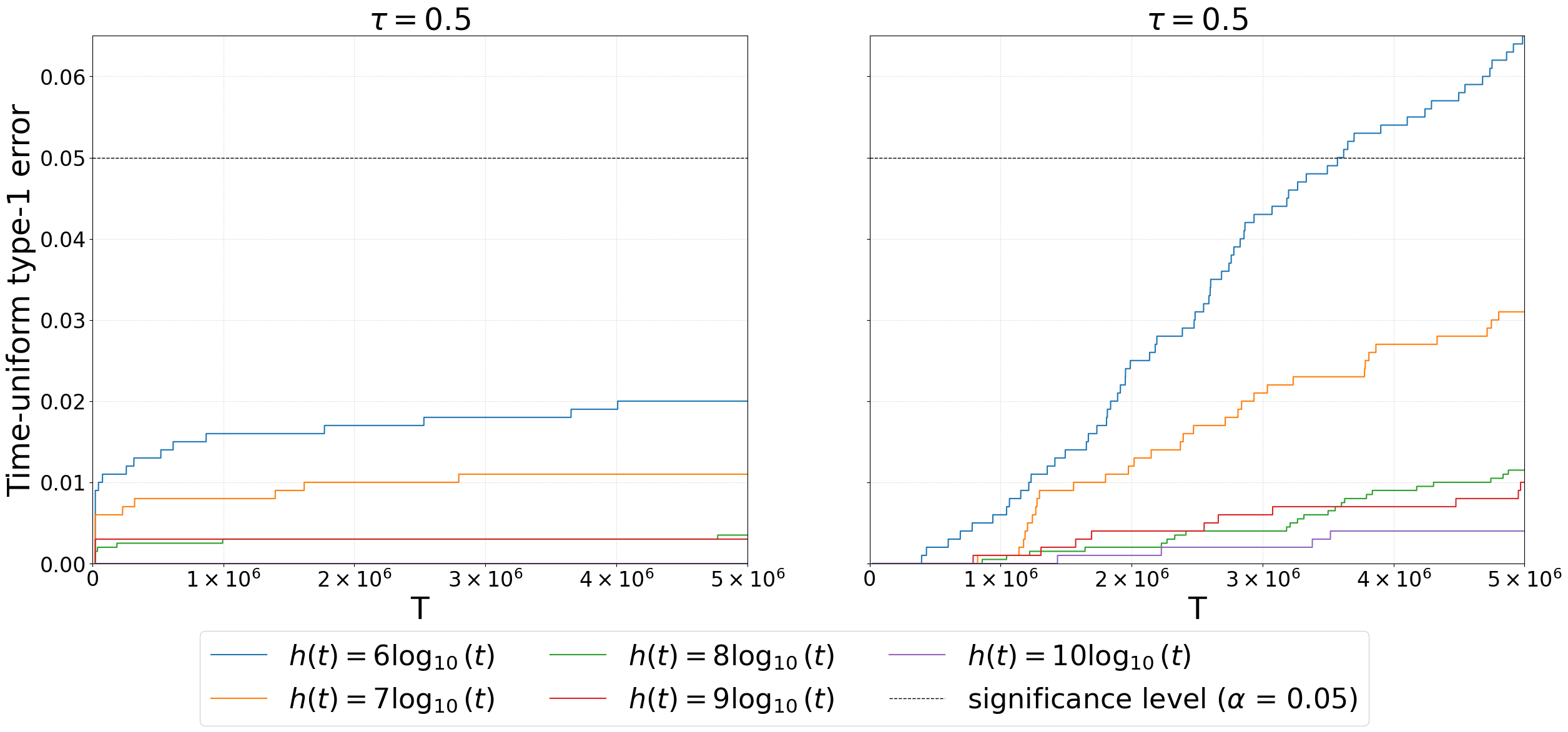}
\caption{Sensitivity to the chain-growth function \(h(t)\).}
\label{fig:sens-ht}
\end{figure}

\begin{figure}[htbp]
\centering
\includegraphics[width=0.90\textwidth]{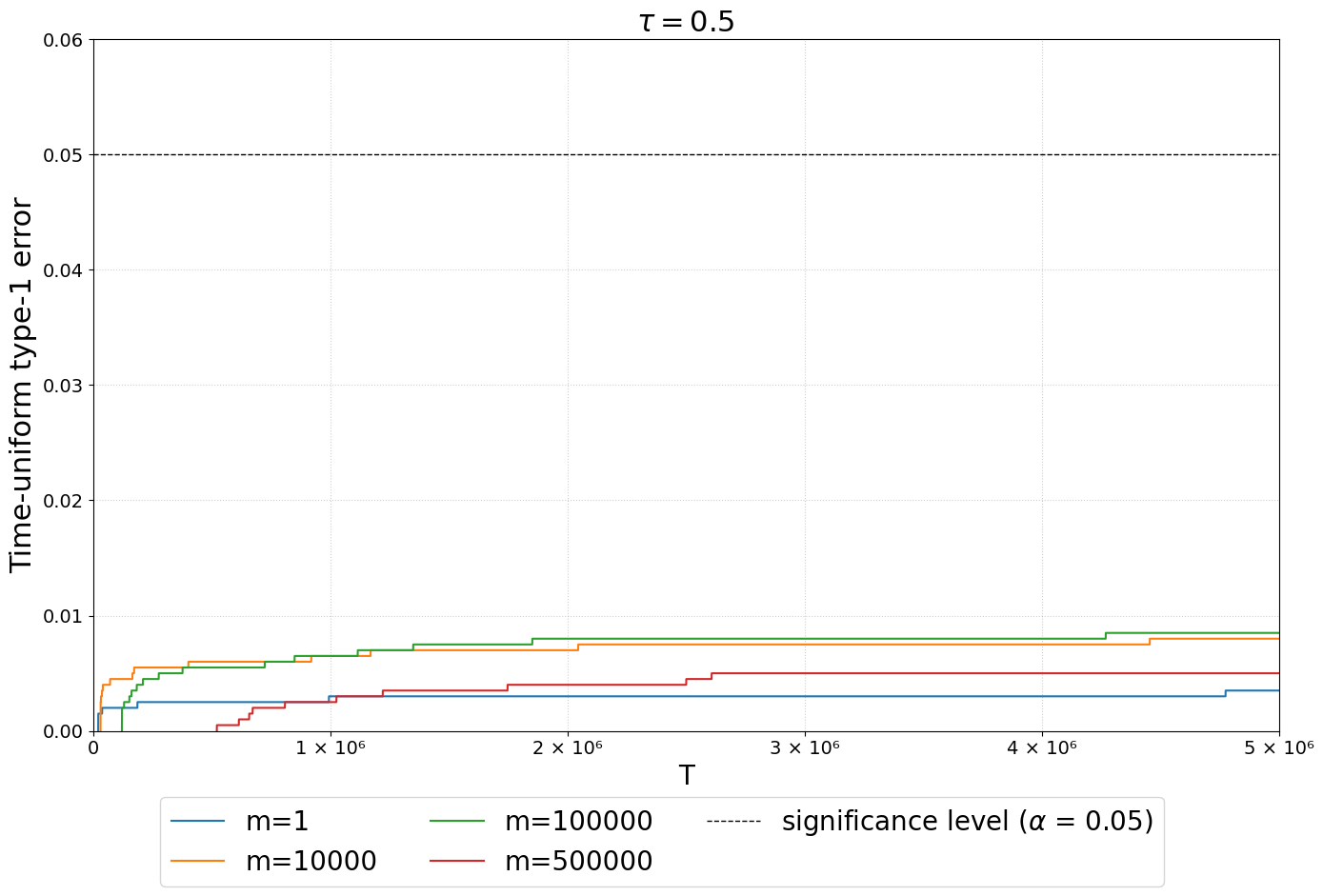}
\caption{Sensitivity to the stitched-boundary start index \(m\).}
\label{fig:sens-m}
\end{figure}

\begin{figure}[htbp]
\centering
\includegraphics[width=0.90\textwidth]{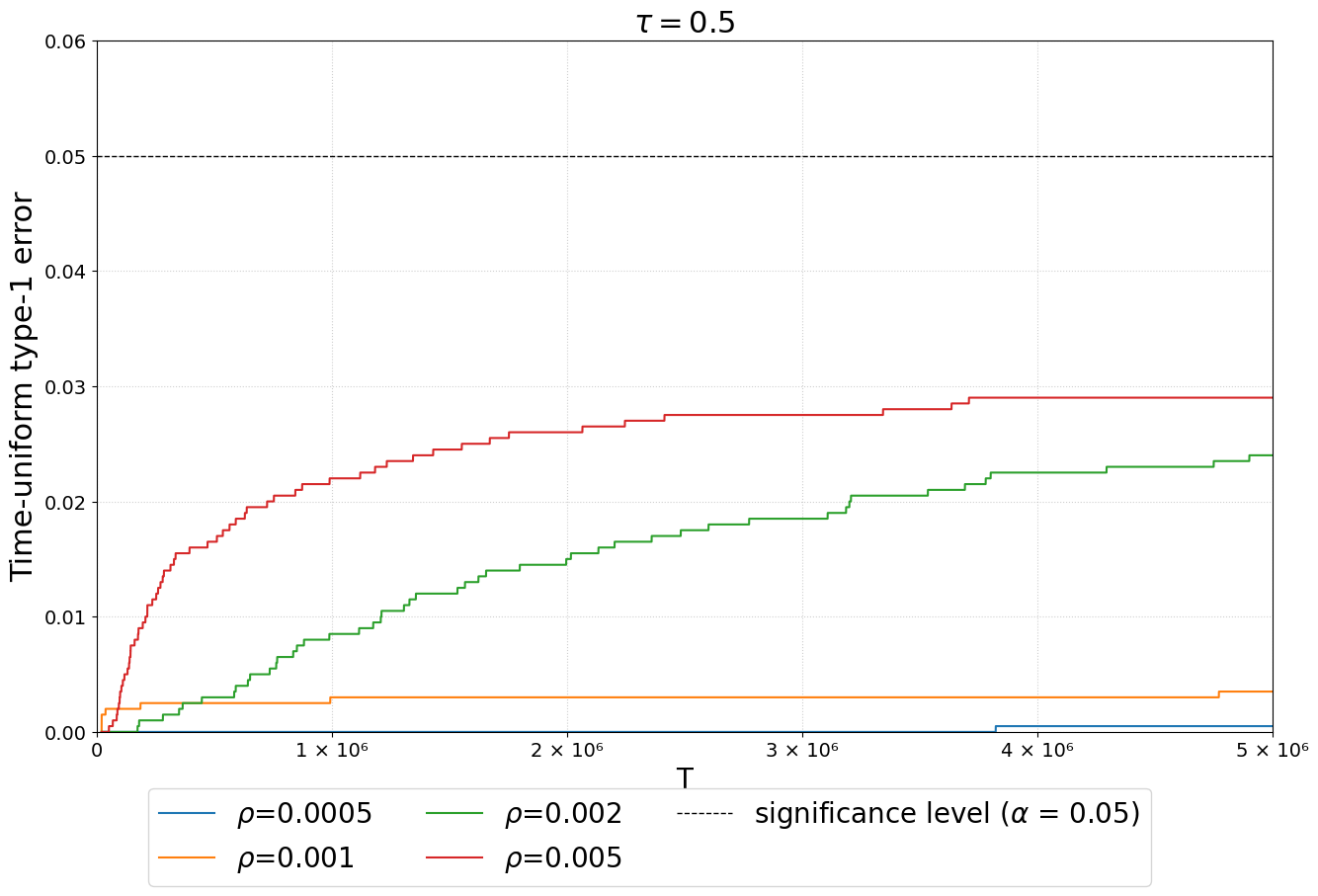}
\caption{Sensitivity to the mixture-boundary parameter \(\rho\).}
\label{fig:sens-rho}
\end{figure}

Figures~\ref{fig:sens-a}--\ref{fig:sens-rho} show broadly stable empirical
anytime error over the tested ranges. The slowest chain-growth schedule,
\(h(t)=6\log_{10}t\), is the notable exception: its Gaussian-mixture curve
slightly exceeds the nominal level. The beta-mixture experiment below therefore
uses \(a=0.6\) and \(h(t)=8\log_{10}t\). The best-arm and A/B experiments use
the polynomial schedule and monitoring start specified in
Section~\ref{subsec:sequential-experiments}.

\begin{figure}[htbp]
\centering
\includegraphics[width=0.90\textwidth]{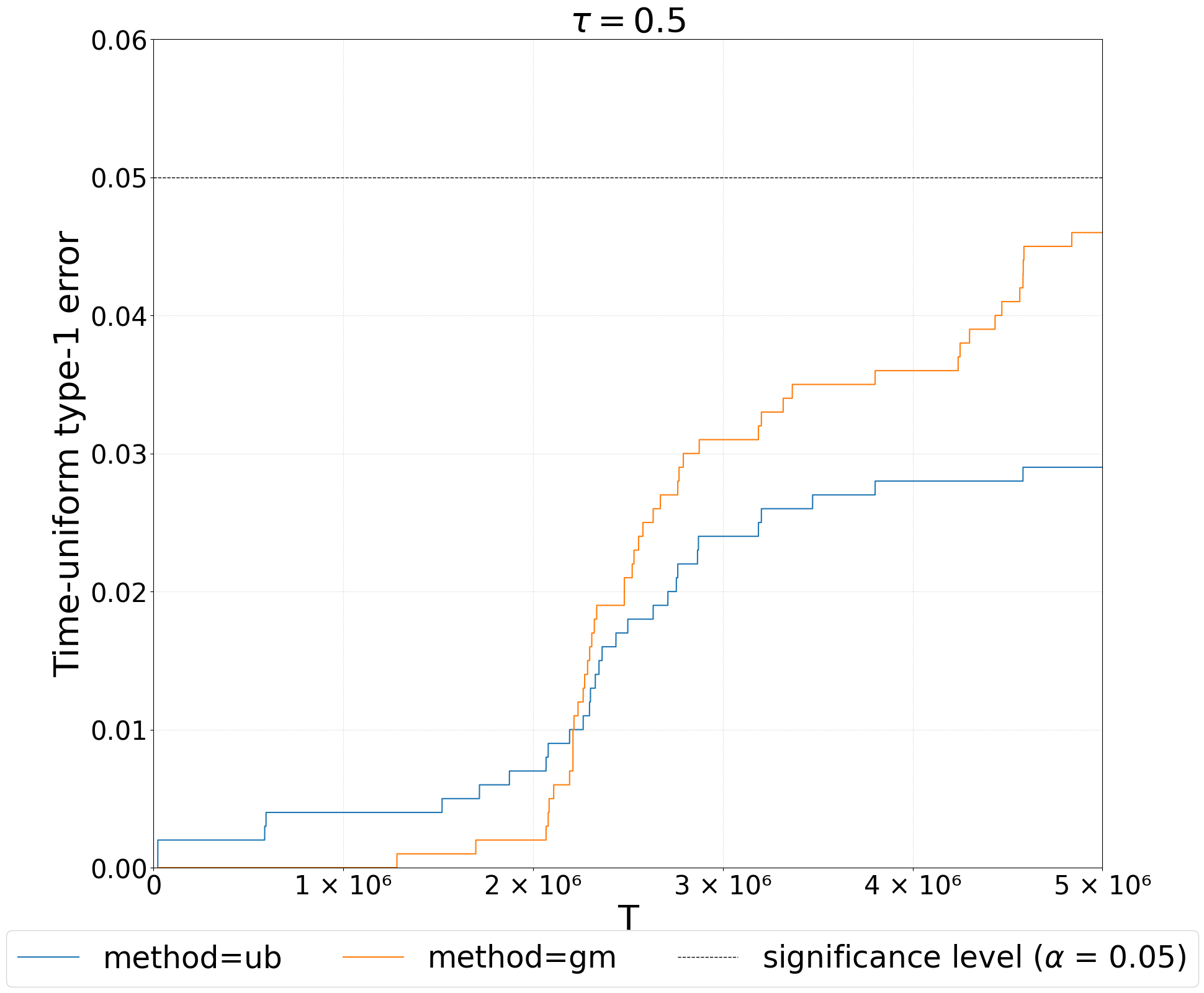}
\caption{Empirical anytime type-I error for the mixture-of-beta stress test with \(\tau=0.5\) and \(r=0.75\).}
\label{fig:mix-beta}
\end{figure}

Finally, we consider a sharply concentrated, multimodal mixture of beta
distributions with density
\[
f(x)=\{\beta_{10,100}(x)+\beta_{100,100}(x)+\beta_{100,10}(x)\}/3,
\]
where \(\beta_{a,b}\) denotes the beta density. The mixture has a globally
Lipschitz density and therefore lies within (A1)--(A2), but its high
curvature and separated modes create a more demanding finite-sample score
geometry than the normal and Cauchy cases. Empirical anytime error remains
controlled in the reported setting \((\tau=0.5,r=0.75)\)
(Figure~\ref{fig:mix-beta}).

\clearpage
\section{Additional Real-Data Analysis}
\label{app:gpa-real-data}

\subsection{Law-School GPA Analysis}\label{app:law-school}
We additionally examine the Law School Admission Council data set, which contains undergraduate GPA and related variables for law-school applicants \citep{wightman1998lsac}. We treat GPA as sensitive educational information. GPA is bounded and comparatively concentrated.

For this data set, we apply a logarithmic transformation before running the LDP stochastic approximation and then back-transform the displayed bounds. The target is the median \((\tau=0.5)\). We use truthful-response rates \(r\in\{0.9,0.8,0.75\}\); the mixture boundary uses \(\rho=0.01\), and the remaining tuning choices match Section~\ref{subsec:simulation-design}. The full-data empirical median, when shown, is a descriptive reference only and is not used by the private procedure.

\begin{figure}[htbp]
\centering
\includegraphics[width=0.90\textwidth]{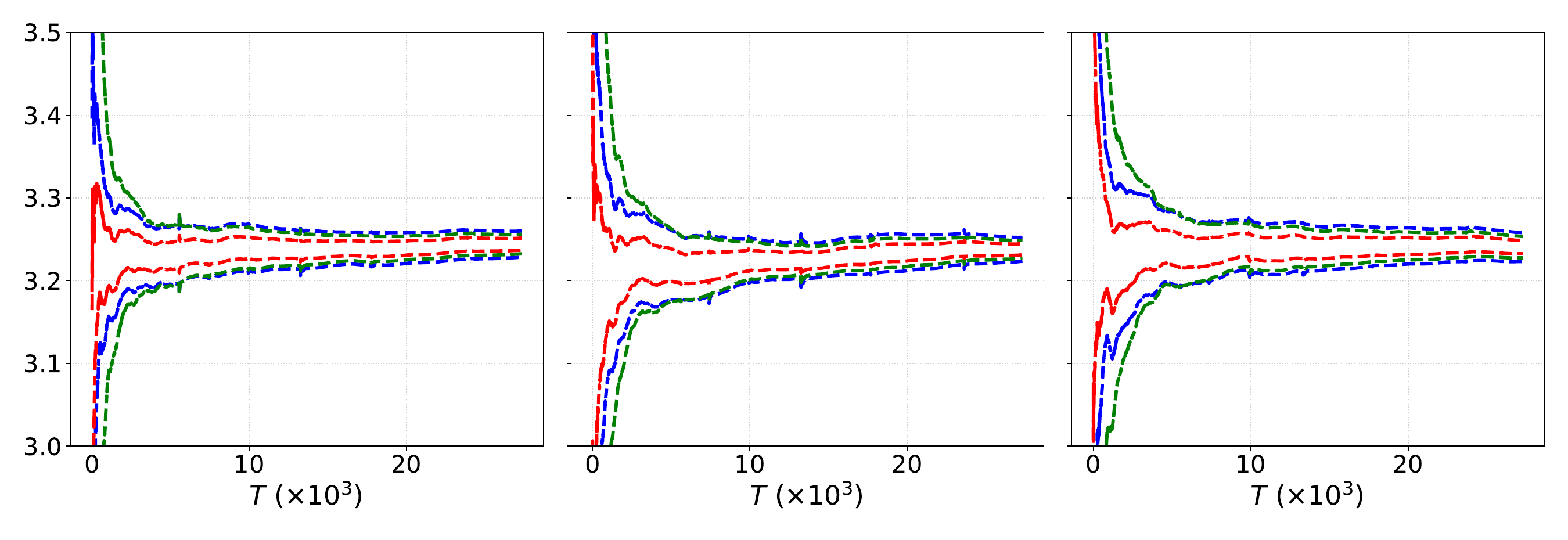}
\caption{Median confidence sequences for GPA in the law-school data. Each panel compares the pointwise interval from Corollary~\ref{cor:pointwise-ci}, the stitched AsympCS and the mixture AsympCS.}
\label{fig:gpa-real-data}
\end{figure}

Figure~\ref{fig:gpa-real-data} reports the resulting confidence sequences. The bands concentrate around a stable central value. Decreasing \(r\) widens the intervals, matching the first-order variance inflation in \eqref{eq:sigma2}. The mixture boundary is shorter over the displayed horizon, while the stitched boundary is more conservative but has the sharper asymptotic order. This real-data result is consistent with the simulation findings and shows that the algorithm can be applied without storing raw individual records at the server.

\clearpage
\subsection{Additional Representative Trajectories for the Salary Applications}
\label{app:salary-trajectories}

For annual salary \(S\) measured in USD, we analyze \(Y=\log(1+S/1000)-4\) and back-transform the displayed bounds to the dollar scale.

\begin{figure}[ht]
\centering
\begin{subfigure}[t]{0.32\textwidth}
\centering
\includegraphics[width=\linewidth]{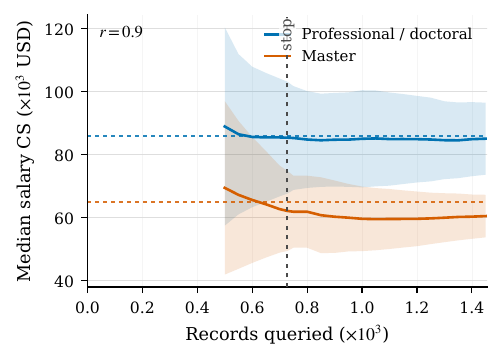}
\caption{\(r=0.9\).}
\end{subfigure}\hfill
\begin{subfigure}[t]{0.32\textwidth}
\centering
\includegraphics[width=\linewidth]{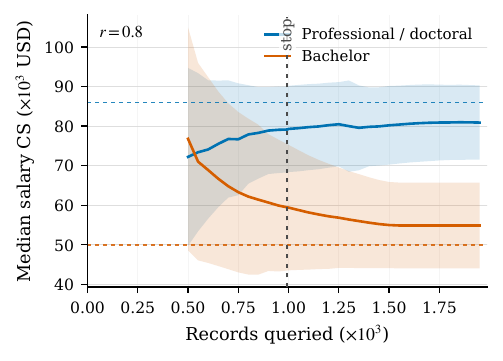}
\caption{\(r=0.8\).}
\end{subfigure}\hfill
\begin{subfigure}[t]{0.32\textwidth}
\centering
\includegraphics[width=\linewidth]{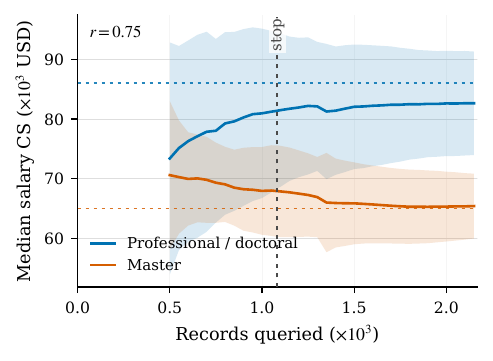}
\caption{\(r=0.75\).}
\end{subfigure}
\caption{Additional representative trajectories for the salary
quantile best-arm application. Each panel shows the back-transformed
median-salary AsympCSs for the selected education group and its strongest
challenger under the indicated truthful-response rate. Horizontal dotted
lines are the corresponding full-data empirical medians, used only as
descriptive references, and the vertical dashed line marks the first
best-arm stopping time.}
\label{fig:salary-bestarm-all-trajectories}
\end{figure}

\begin{figure}[ht]
\centering
\begin{subfigure}[t]{0.32\textwidth}
\centering
\includegraphics[width=\linewidth]{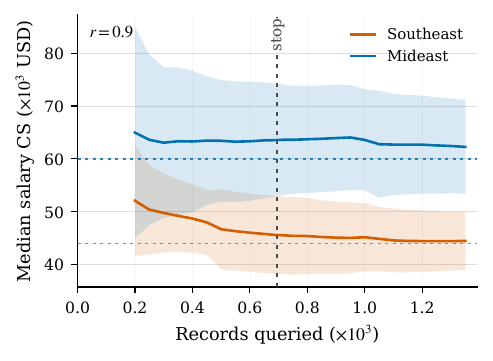}
\caption{\(r=0.9\).}
\end{subfigure}\hfill
\begin{subfigure}[t]{0.32\textwidth}
\centering
\includegraphics[width=\linewidth]{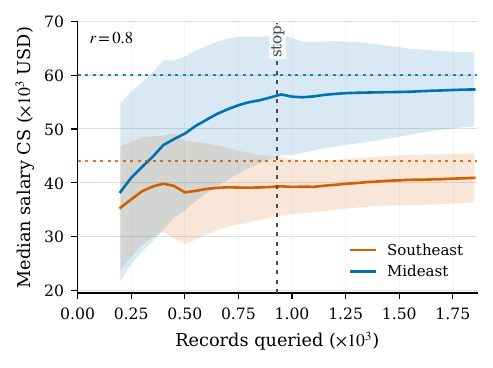}
\caption{\(r=0.8\).}
\end{subfigure}\hfill
\begin{subfigure}[t]{0.32\textwidth}
\centering
\includegraphics[width=\linewidth]{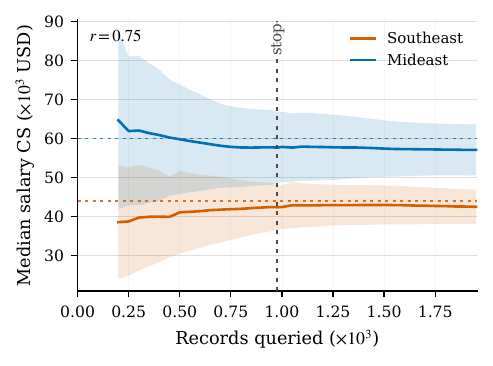}
\caption{\(r=0.75\).}
\end{subfigure}
\caption{Additional representative trajectories for the salary
online quantile A/B test. Each panel shows the back-transformed arm-wise
median-salary AsympCSs for the Southeast and Mideast regions under the
indicated truthful-response rate. Horizontal dotted lines are full-data
empirical medians used only for interpretation, and the vertical dashed line
marks the first time at which the induced median-difference confidence
sequence excludes zero.}
\label{fig:salary-ab-all-trajectories}
\end{figure}

The main text reports compact two-panel summaries of the salary best-arm and
A/B-testing case studies. For completeness, Figures~\ref{fig:salary-bestarm-all-trajectories}
and~\ref{fig:salary-ab-all-trajectories} display the retained representative
arm-wise trajectories separately for each truthful-response rate
\(r\in\{0.9,0.8,0.75\}\). These figures use the same data preprocessing,
dynamic-chain estimators, delayed-start Gaussian-mixture boundaries, and
representative runs as the corresponding main-text analyses.

\clearpage

\bibliography{jmlr_quantile_ldp}

@inproceedings{aamand2025lightweight,
  title = {Lightweight Protocols for Distributed Private Quantile Estimation},
  author = {Aamand, Anders and Boninsegna, Fabrizio and Gentle, Abigail and Imola, Jacob and Pagh, Rasmus},
  booktitle = {Proceedings of the 42nd International Conference on Machine Learning},
  pages = {27--58},
  year = {2025},
  series = {Proceedings of Machine Learning Research},
  volume = {267},
  publisher = {PMLR}
}

@article{alabi2022bounded,
  title = {Bounded Space Differentially Private Quantiles},
  author = {Alabi, Daniel and Ben-Eliezer, Omri and Chaturvedi, Anamay},
  journal = {Transactions on Machine Learning Research},
  year = {2023},
  url = {https://openreview.net/forum?id=sixOD8YVvM}
}

@inproceedings{ben2022archimedes,
  title = {{Archimedes} Meets Privacy: On Privately Estimating Quantiles in High Dimensions under Minimal Assumptions},
  author = {Ben-Eliezer, Omri and Mikulincer, Dan and Zadik, Ilias},
  booktitle = {Advances in Neural Information Processing Systems},
  volume = {35},
  pages = {32450--32464},
  year = {2022}
}

@article{cai2021cost,
  title = {The Cost of Privacy: Optimal Rates of Convergence for Parameter Estimation with Differential Privacy},
  author = {Cai, T. Tony and Wang, Yichen and Zhang, Linjun},
  journal = {The Annals of Statistics},
  volume = {49},
  number = {5},
  pages = {2825--2850},
  year = {2021}
}

@inproceedings{cai2025timeuniform,
  title = {Time-Uniform and Asymptotic Confidence Sequence of Quantile under Local Differential Privacy},
  author = {Cai, Leheng and Hu, Qirui and Sun, Juntao and Wu, Shuyuan},
  booktitle = {Advances in Neural Information Processing Systems},
  volume = {38},
  pages = {114488--114520},
  year = {2025}
}

@article{chen2008nonparametric,
  title = {Nonparametric Estimation of Expected Shortfall},
  author = {Chen, Song Xi},
  journal = {Journal of Financial Econometrics},
  volume = {6},
  number = {1},
  pages = {87--107},
  year = {2008}
}

@article{chen2023recursive,
  title = {Recursive Quantile Estimation: Non-Asymptotic Confidence Bounds},
  author = {Chen, Likai and Keilbar, Georg and Wu, Wei Biao},
  journal = {Journal of Machine Learning Research},
  volume = {24},
  number = {91},
  pages = {1--25},
  year = {2023}
}

@article{chernozhukov2011inference,
  title = {Inference for Extremal Conditional Quantile Models, with an Application to Market and Birthweight Risks},
  author = {Chernozhukov, Victor and Fern{\'a}ndez-Val, Iv{\'a}n},
  journal = {The Review of Economic Studies},
  volume = {78},
  number = {2},
  pages = {559--589},
  year = {2011}
}

@article{draghicescu2009quantile,
  title = {Quantile Curve Estimation and Visualization for Nonstationary Time Series},
  author = {Draghicescu, Dana and Guillas, Serge and Wu, Wei Biao},
  journal = {Journal of Computational and Graphical Statistics},
  volume = {18},
  number = {1},
  pages = {1--20},
  year = {2009}
}

@inproceedings{duchi2013local,
  title = {Local Privacy and Statistical Minimax Rates},
  author = {Duchi, John C. and Jordan, Michael I. and Wainwright, Martin J.},
  booktitle = {Proceedings of the 54th Annual IEEE Symposium on Foundations of Computer Science},
  pages = {429--438},
  year = {2013}
}

@inproceedings{dwork2006our,
  title = {Our Data, Ourselves: Privacy via Distributed Noise Generation},
  author = {Dwork, Cynthia and Kenthapadi, Krishnaram and McSherry, Frank and Mironov, Ilya and Naor, Moni},
  booktitle = {Advances in Cryptology---EUROCRYPT 2006},
  pages = {486--503},
  year = {2006},
  publisher = {Springer}
}

@inproceedings{dwork2009differential,
  title = {Differential Privacy and Robust Statistics},
  author = {Dwork, Cynthia and Lei, Jing},
  booktitle = {Proceedings of the 41st Annual ACM Symposium on Theory of Computing},
  pages = {371--380},
  year = {2009}
}

@article{dwork2014algorithmic,
  title = {The Algorithmic Foundations of Differential Privacy},
  author = {Dwork, Cynthia and Roth, Aaron},
  journal = {Foundations and Trends in Theoretical Computer Science},
  volume = {9},
  number = {3--4},
  pages = {211--407},
  year = {2014}
}

@inproceedings{erlingsson2014rappor,
  title = {{RAPPOR}: Randomized Aggregatable Privacy-Preserving Ordinal Response},
  author = {Erlingsson, {\'U}lfar and Pihur, Vasyl and Korolova, Aleksandra},
  booktitle = {Proceedings of the 2014 ACM SIGSAC Conference on Computer and Communications Security},
  pages = {1054--1067},
  year = {2014}
}

@inproceedings{gillenwater2021differentially,
  title = {Differentially Private Quantiles},
  author = {Gillenwater, Jennifer and Joseph, Matthew and Kulesza, Alex},
  booktitle = {Proceedings of the 38th International Conference on Machine Learning},
  pages = {3713--3722},
  year = {2021},
  series = {Proceedings of Machine Learning Research},
  volume = {139},
  publisher = {PMLR}
}

@article{howard2021time,
  title = {Time-Uniform, Nonparametric, Nonasymptotic Confidence Sequences},
  author = {Howard, Steven R. and Ramdas, Aaditya and McAuliffe, Jon and Sekhon, Jasjeet},
  journal = {The Annals of Statistics},
  volume = {49},
  number = {2},
  pages = {1055--1080},
  year = {2021}
}

@article{howard2022sequential,
  title = {Sequential Estimation of Quantiles with Applications to {A/B} Testing and Best-Arm Identification},
  author = {Howard, Steven R. and Ramdas, Aaditya},
  journal = {Bernoulli},
  volume = {28},
  number = {3},
  pages = {1704--1728},
  year = {2022}
}

@article{hu2022stochastic,
  title = {A Stochastic Approximation Method for Simulation-Based Quantile Optimization},
  author = {Hu, Jiaqiao and Peng, Yijie and Zhang, Gongbo and Zhang, Qi},
  journal = {INFORMS Journal on Computing},
  volume = {34},
  number = {6},
  pages = {2889--2907},
  year = {2022}
}

@inproceedings{liu2023online,
  title = {Online Local Differential Private Quantile Inference via Self-Normalization},
  author = {Liu, Yi and Hu, Qirui and Ding, Lei and Kong, Linglong},
  booktitle = {Proceedings of the 40th International Conference on Machine Learning},
  pages = {21698--21714},
  year = {2023},
  series = {Proceedings of Machine Learning Research},
  volume = {202},
  publisher = {PMLR}
}

@inproceedings{liu2024tuning,
  title = {Tuning-Free Estimation and Inference of Cumulative Distribution Function under Local Differential Privacy},
  author = {Liu, Yi and Hu, Qirui and Kong, Linglong},
  booktitle = {Proceedings of the 41st International Conference on Machine Learning},
  pages = {31147--31164},
  year = {2024},
  series = {Proceedings of Machine Learning Research},
  volume = {235},
  publisher = {PMLR}
}

@article{plevcko2024fairadapt,
  title = {fairadapt: Causal Reasoning for Fair Data Preprocessing},
  author = {Ple{\v c}ko, Drago and Bennett, Nicolas and Meinshausen, Nicolai},
  journal = {Journal of Statistical Software},
  volume = {110},
  number = {4},
  pages = {1--35},
  year = {2024}
}

@article{robbins1970statistical,
  title = {Statistical Methods Related to the Law of the Iterated Logarithm},
  author = {Robbins, Herbert},
  journal = {The Annals of Mathematical Statistics},
  volume = {41},
  number = {5},
  pages = {1397--1409},
  year = {1970}
}

@article{robbins1970boundary,
  title = {Boundary Crossing Probabilities for the {Wiener} Process and Sample Sums},
  author = {Robbins, Herbert and Siegmund, David},
  journal = {The Annals of Mathematical Statistics},
  volume = {41},
  number = {5},
  pages = {1410--1429},
  year = {1970}
}

@inproceedings{smith2011privacy,
  title = {Privacy-Preserving Statistical Estimation with Optimal Convergence Rates},
  author = {Smith, Adam},
  booktitle = {Proceedings of the 43rd Annual ACM Symposium on Theory of Computing},
  pages = {813--822},
  year = {2011}
}

@article{wang2012estimation,
  title = {Estimation of High Conditional Quantiles for Heavy-Tailed Distributions},
  author = {Wang, Huixia Judy and Li, Deyuan and He, Xuming},
  journal = {Journal of the American Statistical Association},
  volume = {107},
  number = {500},
  pages = {1453--1464},
  year = {2012}
}

@article{warner1965randomized,
  title = {Randomized Response: A Survey Technique for Eliminating Evasive Answer Bias},
  author = {Warner, Stanley L.},
  journal = {Journal of the American Statistical Association},
  volume = {60},
  number = {309},
  pages = {63--69},
  year = {1965}
}

@article{waudby2024time,
  title = {Time-Uniform Central Limit Theory and Asymptotic Confidence Sequences},
  author = {Waudby-Smith, Ian and Arbour, David and Sinha, Ritwik and Kennedy, Edward H. and Ramdas, Aaditya},
  journal = {The Annals of Statistics},
  volume = {52},
  number = {6},
  pages = {2613--2640},
  year = {2024}
}

@techreport{wightman1998lsac,
  title = {{LSAC} National Longitudinal Bar Passage Study},
  author = {Wightman, Linda F.},
  institution = {Law School Admission Council},
  year = {1998}
}

@article{xie2024asymptotic,
  title = {Asymptotic Time-Uniform Inference for Parameters in Averaged Stochastic Approximation},
  author = {Xie, Chuhan and Jin, Kaicheng and Liang, Jiadong and Zhang, Zhihua},
  journal = {arXiv preprint arXiv:2410.15057},
  year = {2024}
}

@inproceedings{Lee2022,
  title = {Fast and Robust Online Inference with Stochastic Gradient Descent via Random Scaling},
  author = {Lee, Sokbae and Liao, Yuan and Seo, Myung Hwan and Shin, Youngki},
  booktitle = {Proceedings of the AAAI Conference on Artificial Intelligence},
  volume = {36},
  pages = {7381--7389},
  year = {2022}
}

@inproceedings{anand2019spearphone,
title={{Spearphone}: A Lightweight Speech Privacy Exploit via Accelerometer-Sensed Reverberations from Smartphone Loudspeakers},
author={Anand, S Abhishek and Wang, Chen and Liu, Jian and Saxena, Nitesh and Chen, Yingying},
booktitle={Proceedings of the 14th ACM Conference on Security and Privacy in Wireless and Mobile Networks},
pages={288--299},
doi={10.1145/3448300.3468499},
year={2021}
}

@article{fang2018online,
title={Online bootstrap confidence intervals for the stochastic gradient descent estimator},
author={Fang, Yixin and Xu, Jinfeng and Yang, Lei},
journal={Journal of Machine Learning Research},
volume={19},
number={78},
pages={1--21},
year={2018}
}

@article{hua2016we,
title={We can track you if you take the metro: Tracking metro riders using accelerometers on smartphones},
author={Hua, Jingyu and Shen, Zhenyu and Zhong, Sheng},
journal={IEEE Transactions on Information Forensics and Security},
volume={12},
number={2},
pages={286--297},
year={2017},
publisher={IEEE}
}

@inproceedings{li2022statistical,
title={Statistical Estimation and Online Inference via Local {SGD}},
author={Li, Xiang and Liang, Jiadong and Chang, Xiangyu and Zhang, Zhihua},
booktitle={Proceedings of the 35th Conference on Learning Theory},
pages={1613--1661},
year={2022},
series={Proceedings of Machine Learning Research},
volume={178},
publisher={PMLR}
}

@article{narayanan2006break,
title={How to break anonymity of the {N}etflix prize dataset},
author={Narayanan, Arvind and Shmatikov, Vitaly},
journal={arXiv preprint cs/0610105},
year={2006}
}

@article{su2023higrad,
title={{HiGrad}: Uncertainty Quantification for Online Learning and Stochastic Approximation},
author={Su, Weijie J and Zhu, Yuancheng},
journal={Journal of Machine Learning Research},
volume={24},
number={124},
pages={1--53},
year={2023}
}

@article{zhu2024high,
title={High confidence level inference is almost free using parallel stochastic optimization},
author={Zhu, Wanrong and Lou, Zhipeng and Wei, Ziyang and Wu, Wei Biao},
journal={arXiv preprint arXiv:2401.09346},
year={2024}
}
\end{document}